\documentclass[11pt,a4paper]{article}
\pdfoutput=1

\usepackage{jheppub}
\usepackage{latexsym}
\usepackage{multirow}
\usepackage{color}
\usepackage[usenames,dvipsnames,table]{xcolor}
\usepackage{graphicx}
\usepackage{epsfig}  
\usepackage{dcolumn}
\usepackage{bm}
\usepackage{dcolumn}
\usepackage{textcomp}
\usepackage{float}
\usepackage{subfigure}
\usepackage[]{hyperref}
\usepackage{makecell}
\usepackage{color}
\usepackage{pifont}
\usepackage{url}

 \usepackage{caption}
  \newcommand{\capdef}{}
  \newcommand{\mycaption}[2][\capdef]{\renewcommand{\capdef}{#2}
       \caption[#1]{{\footnotesize #2}}}
  \newcommand{\be}{\begin{equation}}
   \newcommand{\ee}{\end{equation}}

\hypersetup{
  bookmarks=true,         
  unicode=false,          
  pdftoolbar=true,        
 pdfmenubar=true,        
 pdffitwindow=true,     
 pdfstartview={FitH},    
 pdfsubject={Dark matter searches Neutrino Phenomenology},   
 pdfnewwindow=true,      
 pdfcreator={RevTeX},
 colorlinks=true,       
 linkcolor=red,          
 citecolor=blue,        
 filecolor=black,      
 urlcolor=blue,           
  }

\preprint{IP/BBSR/2017-7}
 
\title{Indirect searches of Galactic diffuse dark matter in INO-MagICAL detector}

\author[a,b]{Amina Khatun,}
\author[c,d]{Ranjan Laha,}
\author[a,b]{Sanjib Kumar Agarwalla}

\affiliation[a]{Institute of Physics, Sachivalaya Marg, Sainik School Post, 
Bhubaneswar 751005, India}
\affiliation[b]{Homi Bhabha National Institute, Training School Complex, Anushakti Nagar, Mumbai 400085, India}
\affiliation[c]{Kavli Institute for Particle Astrophysics and Cosmology (KIPAC), Department of Physics, Stanford University, Stanford, CA 94305, USA} 
\affiliation[d]{SLAC National Accelerator Laboratory, Menlo Park, CA 94025, USA}

\emailAdd{amina@iopb.res.in}
\emailAdd{rlaha@stanford.edu}
\emailAdd{sanjib@iopb.res.in}

\abstract{The signatures for the existence of dark matter are revealed only through its
 gravitational interaction. Theoretical arguments support that the Weakly Interacting Massive Particle (WIMP) can be a class of 
 dark matter and it can annihilate and/or decay to Standard Model particles, among which neutrino is a favorable candidate. 
 We show that the proposed 50 kt Magnetized Iron CALorimeter (MagICAL) detector under the India-based Neutrino
 Observatory (INO) project can play an important role in the indirect searches of Galactic diffuse dark matter in the neutrino and antineutrino mode separately.
 We present the sensitivity of 500 kt$\cdot$yr MagICAL detector to set limits on the velocity-averaged 
 self-annihilation cross-section ($\langle\sigma v\rangle$) and decay lifetime ($\tau$) 
 of dark matter having mass in the range of 2 GeV $\leq m_\chi \leq $ 90 GeV and 
 4 GeV $\leq m_\chi \leq $ 180 GeV respectively, assuming no excess over the conventional atmospheric neutrino and antineutrino fluxes at the INO site.  
 Our limits for low mass dark matter constrain the parameter space which has not been explored before. 
We show that MagICAL will be able to set competitive constraints,  
 $\langle\sigma v\rangle\leq 1.87\,\times\,10^{-24}$ cm$^3$ s$^{-1}$ for
 $\chi\chi\rightarrow\nu\bar\nu$ and $\tau\geq 4.8\,\times\,10^{24}$ s for $\chi\rightarrow\nu\bar\nu$ 
 at 90$\%$ C.L. (1 d.o.f.) for $m_\chi$ = 10 GeV assuming the NFW as dark matter density profile. }

\keywords{ICAL, INO, MagICAL, Dark matter, Indirect searches}
\arxivnumber{1703.10221}

\begin{document}

\maketitle

\section{Introduction and Motivation}
\label{sec:Intro}
Plethora of attempts are being made in the intensity, energy, and cosmic frontiers
to build up knowledge about the Universe. Recent observations by Planck satellite\,\cite{Ade:2015xua} confirm
that  the baryonic and unknown non-baryonic matter (dark matter) contribute $\sim$\,4.8$\%$ and $\sim$\,26$\%$ of the total energy density of the Universe respectively. 
The first indication for the existence of dark matter (DM) in the Universe was made by the Swiss astronomer Fritz Zwicky\,\cite{Zwicky:1933gu}. This 
observation was put on a solid footing by Vera Rubin and her collaborators\,\cite{Rubin:1970zza}.
The astrophysical\,\cite{Strigari:2013iaa,Clowe:2006eq} and cosmological observations\,\cite{Komatsu:2014ioa,Steigman:2012ve} 
confirm the existence of dark matter from the length scales of a few kpc to a few Gpc.  
 
All the astrophysical evidences of dark matter are through its gravitational interactions.
The non-gravitational particle physics properties of DM particles are completely unknown. 
The relic abundance of cold dark matter (CDM) in the Universe is matched assuming a $\sim$100 GeV 
dark matter particle with electro-weak coupling strength.
This class of particles is known as Weakly Interacting Massive Particle (WIMP)\,\cite{Jungman:1995df,Bertone:2004pz,Bergstrom:2000pn}.
Supersymmetry, one of the most favored beyond-the-Standard Model theory, also predicts more than one dark
matter candidates including the WIMP\,\cite{Ellis:2010kf}.

There are three types of detection methods for the search of DM: \textbf{(i) Direct detection:} 
DM particles are detected by observing recoiled nuclei from the scattering of DM particles in the laboratory. 
Experiments such as DAMA/LIBRA\,\cite{Bernabei:2010mq}, LUX\,\cite{Akerib:2015rjg}, CDMS\,\cite{Agnese:2015nto}, XENON\,\cite{Aprile:2012nq},
DarkSide\,\cite{Agnes:2015ftt}, and PandaX\,\cite{Xiao:2015psa} pursue this strategy. 
\textbf{(ii) Indirect detection:} It is possible that DM particles can decay and/or annihilate to any of the Standard Model
(SM) particles like $\nu\bar\nu$, $t\bar{t}$, $b\bar{b}$ etc. 
An excess (over standard astrophysical backgrounds) of these SM particles can be searched for to understand dark matter. 
The unstable SM particles decay to produce neutrinos and photons which can be searched for indirect detection.
The prospects of dark matter searches through neutrino portal
has been studied in the literature\,\cite{Lindner:2010rr,Agarwalla:2011yy,Farzan:2011ck,Mijakowski:2011zz,Blennow:2013pya,Gustafsson:2013gca,
Aisati:2015ova,Anchordoqui:2015lqa,Arina:2015zoa,Macias:2015cna,Gonzalez-Macias:2016vxy,Zornoza:2016ggm,Garcia-Cely:2017oco,Beacom:2006tt,Yuksel:2007ac}. 
Fermi-LAT presents the analysis of its collected data of gamma 
rays having the energy in the range of 200 MeV to 500 GeV from Galactic halo in 5.8 years in Ref.\,\cite{Ackermann:2015lka}. 
Multiwavelength searches for dark matter have complementary reach\,\cite{Laha:2012fg}. Our focus in this work
is indirect detection of dark matter via neutrinos and antineutrinos.
\textbf{(iii) Collider searches:} The searches for supersymmetric DM candidates are carried out in LHC\,\cite{Khachatryan:2015bbl, Chatrchyan:2012tea,Aad:2012fw}.

The 50 kt Magnetized Iron CALorimeter (MagICAL\footnote{The ``MagICAL" name is used here as the abbreviation of 
 Magnetized Iron CALorimeter which is commonly known as ICAL detector. We prefer the name MagICAL to 
 emphasize that magnetic field is present in the ICAL detector, which enable us to separate neutrino and anti-neutrino events.}) 
 detector is proposed to  be built by the India-based Neutrino Observatory (INO) project
to observe the atmospheric neutrino and antineutrino separately having energy in multi-GeV range and covering a wide 
ranges of path lengths  (few km to few thousands of km) through the Earth matter. The primary mission of the 
MagICAL detector is to unravel the mass ordering\footnote{Two 
distinct patterns of neutrino masses are allowed: $m^2_3>m^2_2>m^2_1$, known as normal ordering (NO) where $\Delta m^2_{31}\,(\equiv m^2_3 - m^2_1)>0$ and 
$m^2_2>m^2_1>m^2_3$, called inverted ordering (IO) where $\Delta m^2_{31}<0$.} (MO) of 
neutrino using the Earth matter effect\,\cite{Ghosh:2012px,Devi:2014yaa,Ahmed:2015jtv,Mohan:2016gxm} and to
measure the neutrino mixing parameters precisely\,\cite{Thakore:2013xqa,Kaur:2014rxa,Devi:2014yaa}. The MagICAL detector has also the potential 
to explore the physics beyond the Standard Model\,\cite{Dash:2014fba,Chatterjee:2014oda,Chatterjee:2014gxa,Choubey:2015xha,Behera:2016kwr}. 
In our study, we show that the MagICAL detector can play a very important role in the indirect search of
DM having mass in the multi-GeV range with the help of its excellent detection efficiency, energy, and angular resolutions.
We explore the sensitivity of the MagICAL detector to detect the neutrino and antineutrino events coming from the diffuse dark matter annihilation/decay
in the Milky Way galaxy. We present the constraint on the self-annihilation cross-section ($\langle \sigma v \rangle$) and the decay lifetime ($\tau$) of diffuse
dark matter having mass in the range [2, 90] GeV and [4, 180] GeV respectively using 500 kt$\cdot$yr exposure of the MagICAL detector. 

We describe the dark matter density profile and the calculation of annihilation 
and decay rate of dark matter in section\,\ref{sec:input}. The key features of the MagICAL detector is presented in section\,\ref{sec:key features}. 
Section\,\ref{sec:event spectrum} deals with the expected  
event distribution of atmospheric and DM induced neutrinos in the MagICAL detector. We present the simulation method 
in section\,\ref{sec:simulation}. The prospective limits on the self-annihilation cross-section
and decay lifetime of dark matter are presented in section\,\ref{sec:results}. We compare our results with the existing bounds from other experiments.
We also study the flux upper limit due to dark matter induced neutrinos in the MagICAL detector.
We conclude in section\,\ref{sec:conclusions}.

\section{Discussions on dark matter}
\label{sec:input}
\subsection{Dark matter density profile}
The general parameterization of a spherically symmetric dark matter density profile is given by
\begin{equation}
 \rho(r) = \frac{\rho_{0}}{\big[\delta\,+\,r/r_{s}\big]^\gamma\,.\,\big[1+(r/r_{s})^\alpha\big]^{(\beta-\gamma)/\alpha}} \,.
 \label{eq1}
\end{equation}
The density, $\rho(r)$, is expressed in GeV cm$^{-3}$ and $\it{r}$ is the distance from the center of the galaxy in kpc. 
The parameter, $r_{s}$, is the scale radius in kpc. The shape of the outer profile is controlled by $\alpha$ and $\beta$, 
whereas $\gamma$ parametrizes the slope of the inner profile. 
The dark matter density at the Solar radius (R$_{sc}$) is denoted by $\rho_{sc}$. We assume R$_{sc}$ = 8.5 kpc\,\cite{2016arXiv161002457D}. 
The normalization constant, $\rho_{0}$, and all the results are calculated using the values of parameters as given in table\,\ref{table0}. 

Numerical simulations which involve only dark matter particles predict a cuspy profile\,\cite{Navarro:1995iw,Diemand:2006ik,Stadel:2008pn,Navarro:2008kc}. 
Although these simulations reproduce the large-scale structure of the Universe, yet this prescription has challenges at scales below the size of a typical galaxy.
It has been proposed that the addition of baryons can solve all of these small scale issues, although the results vary\,\cite{Stinson:2012uh,DiCintio:2013qxa,
Tollet:2015gqa,Chan:2015tna,Marinacci:2013mha,Kim:2013jpa,Schaye:2014tpa,Schaller:2014uwa,Sawala:2015cdf}. 
Present observations are not yet precise enough to distinguish between a cored and a cuspy profile\,\cite{Cerdeno:2016znc}. 

\begin{figure}[h]
\subfigure[]{\includegraphics[width=8.2cm]{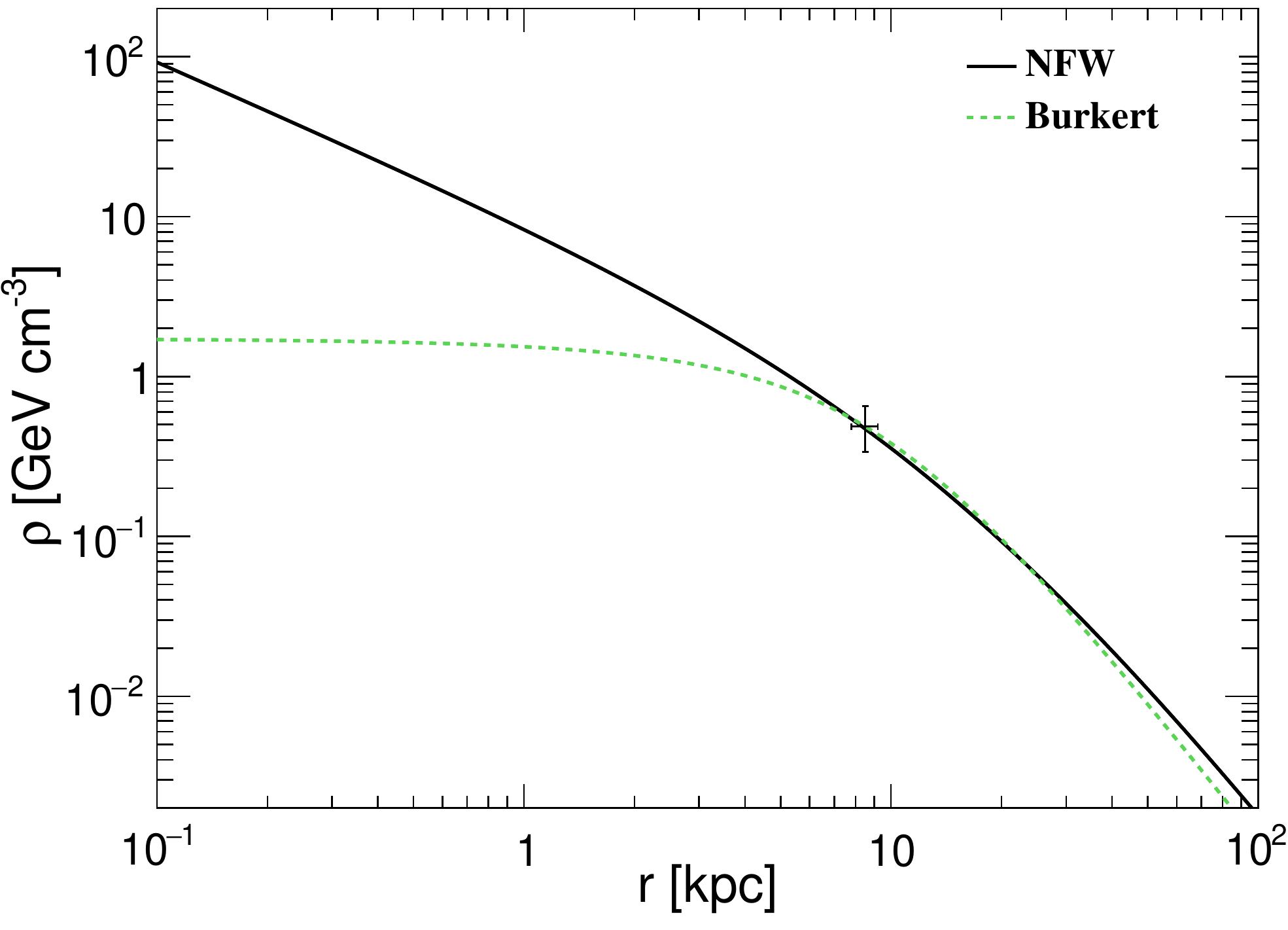}}
\subfigure[]{\includegraphics[width=6.5cm]{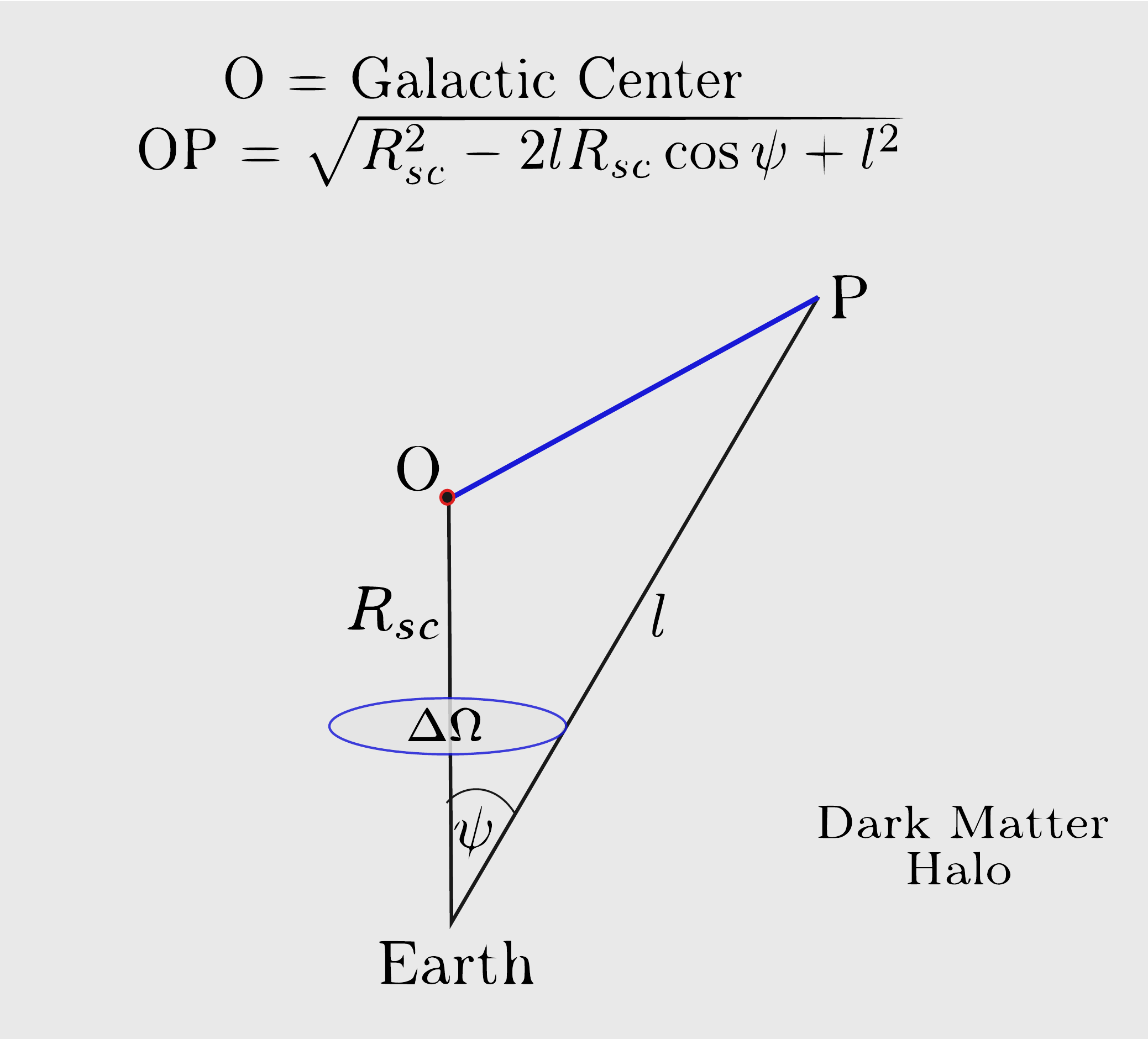}} 
\mycaption{(a) Distribution of the dark matter density in the Milky Way galaxy for the NFW (black solid line) and Burkert profiles (green dashed line). 
The observational bounds on local dark matter density ($\rho_{sc}$) and the solar radius ($R_{sc}$) and 
their 2$\sigma$ uncertainties are indicated\,\cite{Strigari:2013iaa,Aartsen:2015xej}.
(b) A schematic diagram of some part of the Milky Way dark matter halo. The Galactic center (GC) is denoted by O 
and $R_{sc}$ is the distance between the Earth and the GC.  The parameter
$l$ is the distance between point P and the Earth. The angle made at the Earth by points P and O and 
the corresponding solid angle are denoted by $\psi$ and $\Delta\Omega$ respectively.}
\label{fig1}
\end{figure}

To take this DM halo uncertainty into account, we generate all 
the results with two different DM profiles: Navarro Frenk White
(NFW) profile \,\cite{Navarro:1995iw}, which represents cuspy halos, and the Burkert profile\,\cite{1999dmap.conf..375B}, which represents 
 cored halos. The values of different parameters associated with these profiles 
are taken from Ref.\,\cite{Aartsen:2015xej}. In Fig.\,\ref{fig1}(a), we plot the NFW and Burkert dark matter density profiles 
with distance $r$\, from the center of the Milky Way galaxy by the black solid 
and green dashed lines respectively.

For conservativeness, we do not consider the effects of dark matter substructure.  Depending on the 
value of the minimum halo mass and other astrophysical uncertainties, this can give a substantial contribution to the signal discussed 
here\,\cite{Ando:2005hr,Ng:2013xha,Campbell:2014bca,Correa:2015dva,Bartels:2015uba,Moline:2016pbm}.

\begin{table}[htb!]
 \centering
\begin{tabular}{|c|c|c|c|}
\hline
 & ($\alpha, \beta, \gamma, \delta$ ) & $\rho_{sc}$ [GeV cm$^{-3}$] & $r_{s}$ [kpc]\\
\hline
NFW & (1, 3, 1, 0) &  0.471 &  16.1 \cr 
Burkert & (2, 3, 1, 1) & 0.487 & 9.26  \cr
\hline
\end{tabular}
\mycaption{The value of parameters associated with the NFW and Burkert profiles 
are listed here. They are same as in Ref.\,\cite{Aartsen:2015xej}.}
\label{table0}
\end{table}
In Fig.\,\ref{fig1}(b), a schematic diagram of a small portion of
the Milky Way DM halo is shown with O as the Galactic center (GC). The dark matter density at point P with its distance $l$ from 
the Earth is a function of the length OP = $\sqrt{R_{sc}^{2} - 2 l R_{sc} \cos\psi +l^2}$.  
The angle made at the Earth by points P and O is $\psi$ and the corresponding solid angle is
$\Delta\Omega$. 
\subsection{Annihilation of dark matter}
We consider the annihilation between a dark matter particle ($\chi$)
and its antiparticle ($\bar\chi$) to produce a neutrino and an antineutrino in the final state with 
100$\%$ branching ratio:
\begin{equation}
 \chi + \bar\chi \rightarrow \nu + \bar\nu \,.
 \label{eq2}
\end{equation}
The neutrinos and antineutrinos of $e$, $\mu$, and $\tau$ flavors are assumed to be produced in 1:1:1 ratio at source.
This ratio remains same on arrival at the Earth surface due to loss of coherence 
while propagating through astrophysical distances (see appendix~\ref{app1}).

The number of $\nu/ \bar{\nu}$ from a direction $\psi$ due to the annihilation of dark matter 
particles is proportional to the line of sight integration of the square of dark matter density: 
\begin{equation}
\mathcal{J}^{ann}(\psi) = \frac{1}{R_{sc}\rho^{2}_{sc}} \int_0^{l_{max}} dl\,\, \rho^2 \left(\sqrt{R_{sc}^{2} - 2 l R_{sc} \cos\psi +l^2} \right)\,.
 \label{eq3}
\end{equation}

The factor $\frac{1}{R_{sc}\rho^{2}_{sc}}$ is included to make $\mathcal{J}^{ann}(\psi)$ dimensionless. 
The upper limit $l_{max}$ is the distance between the observer and the farthest point (denoted by P$'$) in the Milky Way
halo at the angle $\psi$.  The radius of
Milky Way galaxy is $R_{\rm MW}$ (= OP$'$ = 100 kpc), and thus 
\begin{equation}
 l_{max} = \sqrt{\big(R^2_{MW} - R^2_{sc} \sin^2\psi\big) } + R_{sc}\cos\psi \,.
\label{eq4}
 \end{equation}
Increase of $R_{\rm MW}$ to 150 kpc enhances the value of $\mathcal{J}^{ann}(\psi = 180^{\circ})$ by 0.03$\%$. The average value of $\mathcal{J}^{ann}(\psi) $
 over a solid angle 2$\pi \int^{\psi}_0 \sin\psi^{\prime} d\psi^{\prime} = 2\pi(1-\cos\psi) $ is 
\begin{equation}
 \mathcal{J}_{\Delta\Omega}^{ann}(\psi) = \frac{1}{2\pi(1-\cos\psi)} \int^1_{\cos\psi} 2\pi \,d(\cos\psi^{\prime}) \,\mathcal{J}^{ann}(\psi^{\prime})\,.
 \label{eq5}
\end{equation}\\
\begin{figure}[htb!]
\subfigure{\includegraphics[width = 7.5cm]{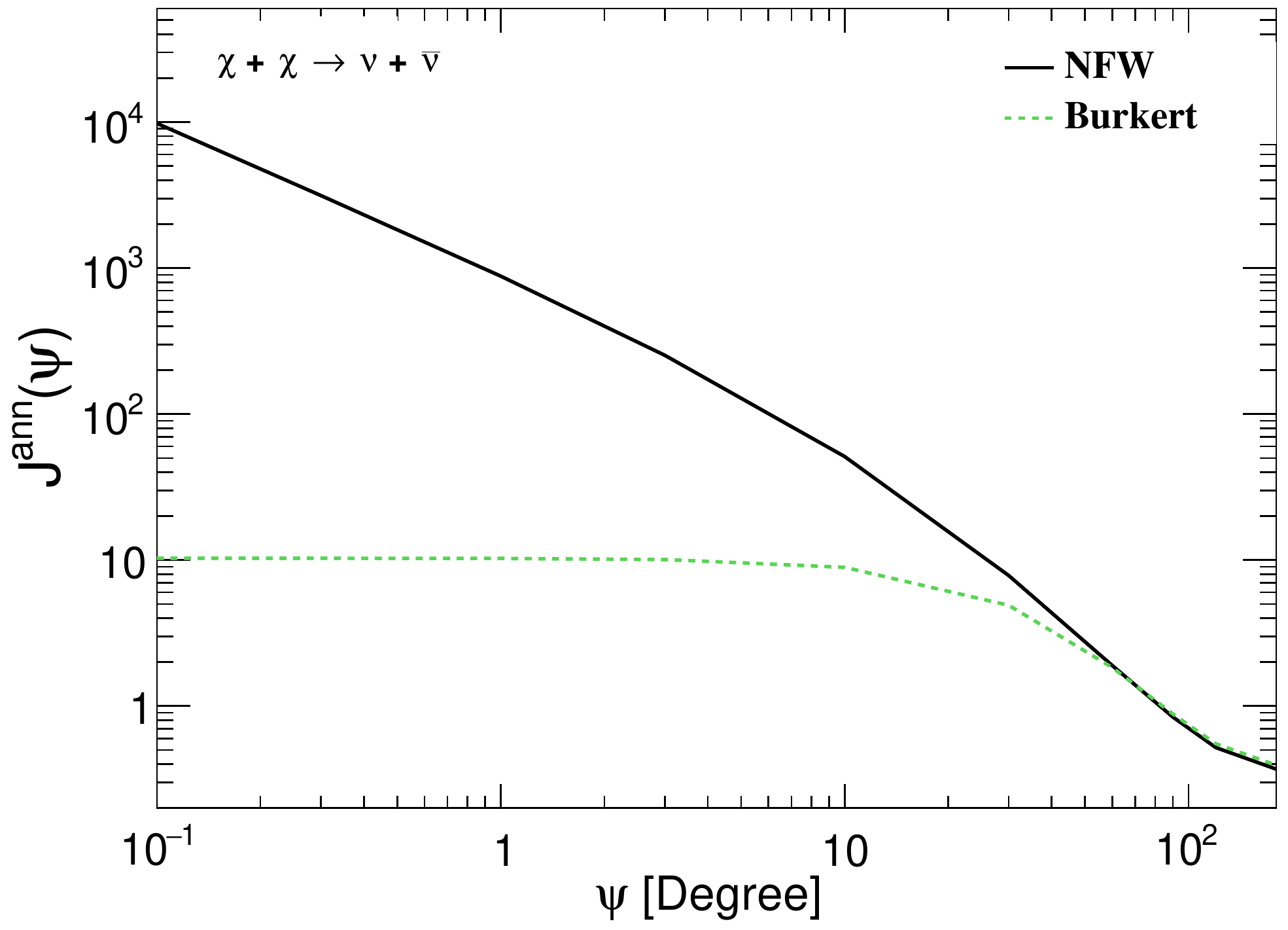}}
\subfigure{\includegraphics[width = 7.5 cm]{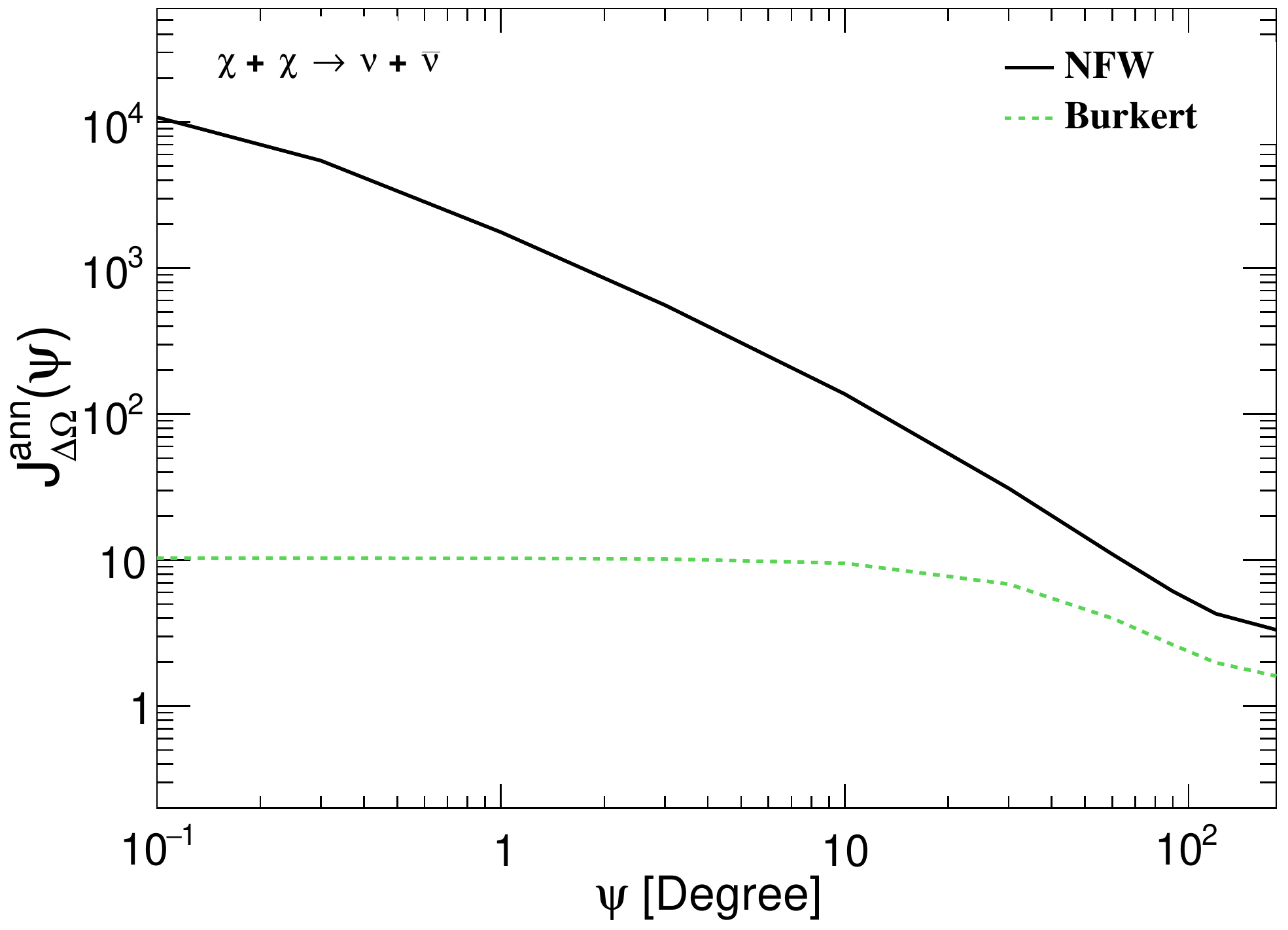}}
\mycaption{The value of $\mathcal{J}^{ann}(\psi)$ (see Eq.\,\ref{eq3}) and its average ($\mathcal{J}_{\Delta\Omega}^{ann}(\psi)$) 
over solid angle $\Delta\Omega$ = $2\pi(1-\cos\psi)$ (see Eq.\,\ref{eq5}) 
are shown in left and right panels. In both the panels, black solid and green dashed lines present the corresponding quantities 
for the NFW and Burkert profiles respectively. We use $\mathcal{J}^{ann}_{\Delta\Omega}(\psi\,=\,180^\circ)$ for the diffuse 
dark matter analysis, which has values 3.33 and 1.6 for the NFW and Burkert profiles respectively.}
\label{fig2}
\end{figure}
The variation of $\mathcal{J}^{ann}(\psi)$ and $\mathcal{J}_{\Delta\Omega}^{ann}(\psi)$ with angle $\psi$ are 
shown by the black solid (green dashed) lines in left and right panels of Fig.\,\ref{fig2} respectively using the NFW (Burkert) DM halo profile. 
The value of $\mathcal{J}^{ann}_{\Delta\Omega}$ = 3.33 for the NFW profile and $\mathcal{J}^{ann}_{\Delta\Omega}$ = 1.6 for the Burkert
profile with $\Delta\Omega$ = 4$\pi$.
The flux of each flavor of $\nu/ \bar{\nu}$ per unit energy range per unit solid angle (in units of GeV$^{-1}$\,sr$^{-1}$\,cm$^{-2}$\,s$^{-1}$)
produced in the final state of dark matter particles annihilation is given by
\begin{equation}
 \frac{d^2\Phi^{ann}_{\nu/\bar\nu}}{{dE}\,d\Omega} = \frac{\langle \sigma_A v\rangle}{2} \mathcal{J}^{ann}_{\Delta\Omega}\,
 \frac{R_{sc}\rho^2_{sc}}{4\pi\,m_{\chi}^2}\,\,\frac{1}{3} \frac{{dN^{ann}}}{{dE}}\,,
 \label{eq6}
\end{equation}
where $\langle \sigma_A v\rangle$ is the self-annihilation cross-section in units of cm$^3$\,s$^{-1}$. 
The factor $\frac{1}{2}$ is included as we assume the dark matter particle is same as its 
own antiparticle. The factor $\frac{1}{3}$ takes into account the flavor ratio of 
$\nu/ \bar\nu$ on the Earth's surface. The probability of $\nu_{e}$, $\nu_\mu$, and $\nu_\tau$ 
to be produced in the final state are the same. Therefore the flux of $\nu$/$\bar\nu$ with 
each lepton flavor is calculated as the total $\nu$/$\bar\nu$ flux divided by the total number 
of lepton generations, which gives rise to the $\frac{1}{3}$ factor in Eq.\,\ref{eq6}. 
The factor 4$\pi$ in the denominator is for the isotropic production of $\nu\bar\nu$ in annihilation of dark matter.
The parameter $m_{\chi}$ is mass of the DM particles in units of GeV.
The energy spectrum of $\nu/ \bar{\nu}$ is given by  
\begin{equation}
 \frac{{dN^{ann}}}{{dE}} = \delta(E_{\nu/\bar\nu} - m_\chi)\,,
 \label{eq7}
\end{equation}
since dark matter particles in our galaxy are non-relativistic (local velocity $\sim$ 10$^{-3}$c).
\subsection{Decay of dark matter}
A dark matter particle is assumed to decay into $\nu_{e}+\bar{\nu}_{e}$, 
$\nu_\mu+\bar{\nu}_\mu$, and $\nu_\tau+\bar{\nu}_{\tau}$ with equal branching ratio:
\begin{equation}
 \chi \rightarrow \nu + \bar\nu\,.
 \label{eq7p1}
\end{equation}
The $\nu/\bar{\nu}$ flux from dark matter decay is proportional to the line of sight integral of the dark matter distribution, $\mathcal{J}^{dec}(\psi)$, with
\begin{equation}
\mathcal{J}^{dec}(\psi) = \frac{1}{R_{sc}\rho_{sc}} \int_0^{l_{max}} dl\,\, \rho \left(\sqrt{R_{sc}^{2} - 2 l R_{sc} \cos\psi +l^2} \right)\,.
\label{eq8}
\end{equation}
The quantity $R_{sc}\rho_{sc}$ in the denominator makes $\mathcal{J}^{dec}(\psi)$ 
dimensionless.  All other symbols have same meaning as before.  The quantity $\mathcal{J}_{\Delta\Omega}^{dec}(\psi)$ represents the average value of
$\mathcal{J}^{dec} (\psi)$ over the solid angle $\Delta\Omega\,=\,2\pi(1-\cos\psi)$:
\begin{equation}
 \mathcal{J}_{\Delta\Omega}^{dec}(\psi) = \frac{1}{2\pi(1-\cos\psi)} \int^1_{\cos\psi} 2\pi\, {d}(\cos\psi^{\prime})\, 
 \mathcal{J}^{dec}(\psi^{\prime})\,.
 \label{eq9}
\end{equation}
\begin{figure}[htb!]
\subfigure{ \includegraphics[width = 7.5cm]{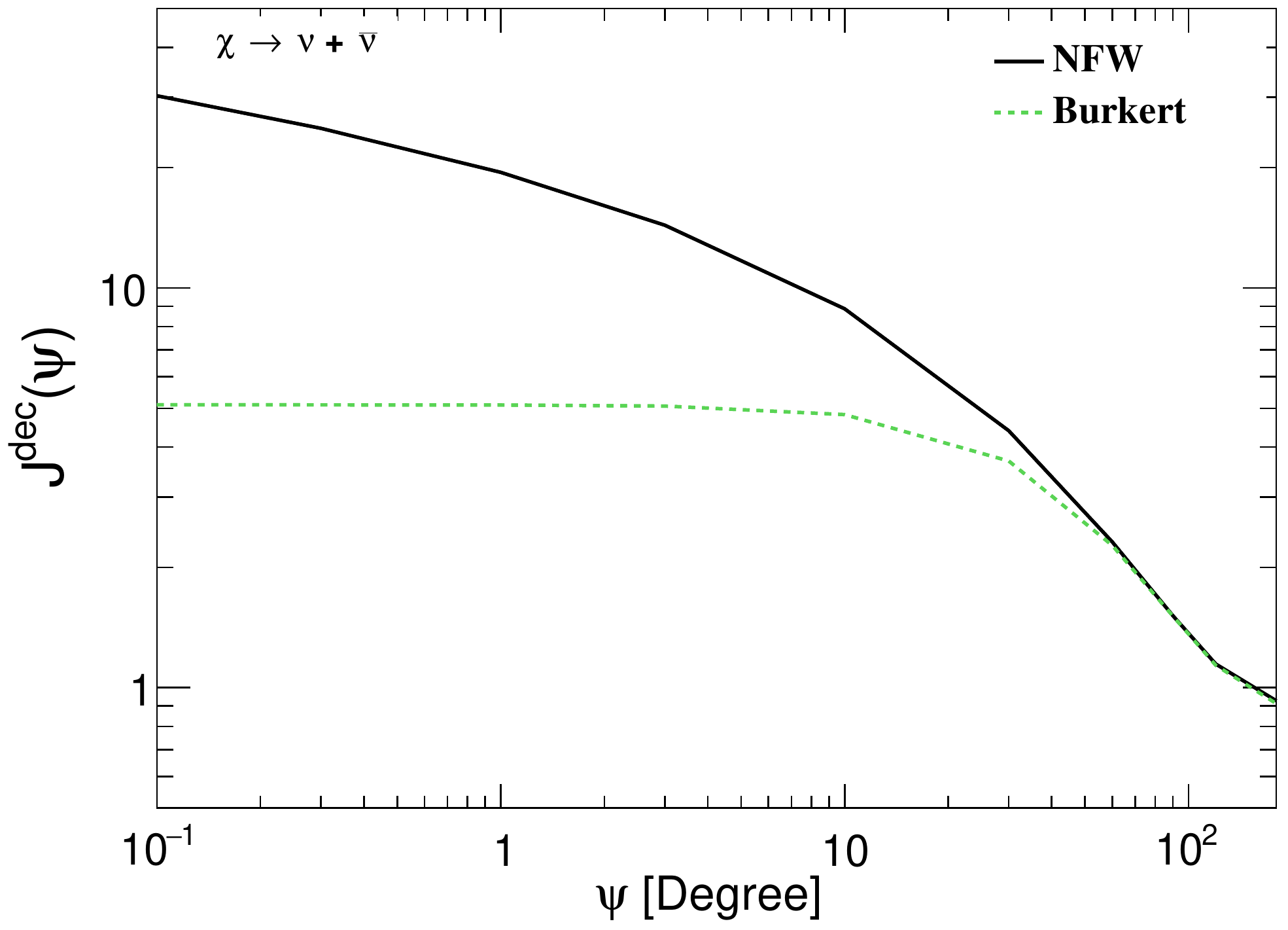}}
\subfigure{\includegraphics[width = 7.5cm]{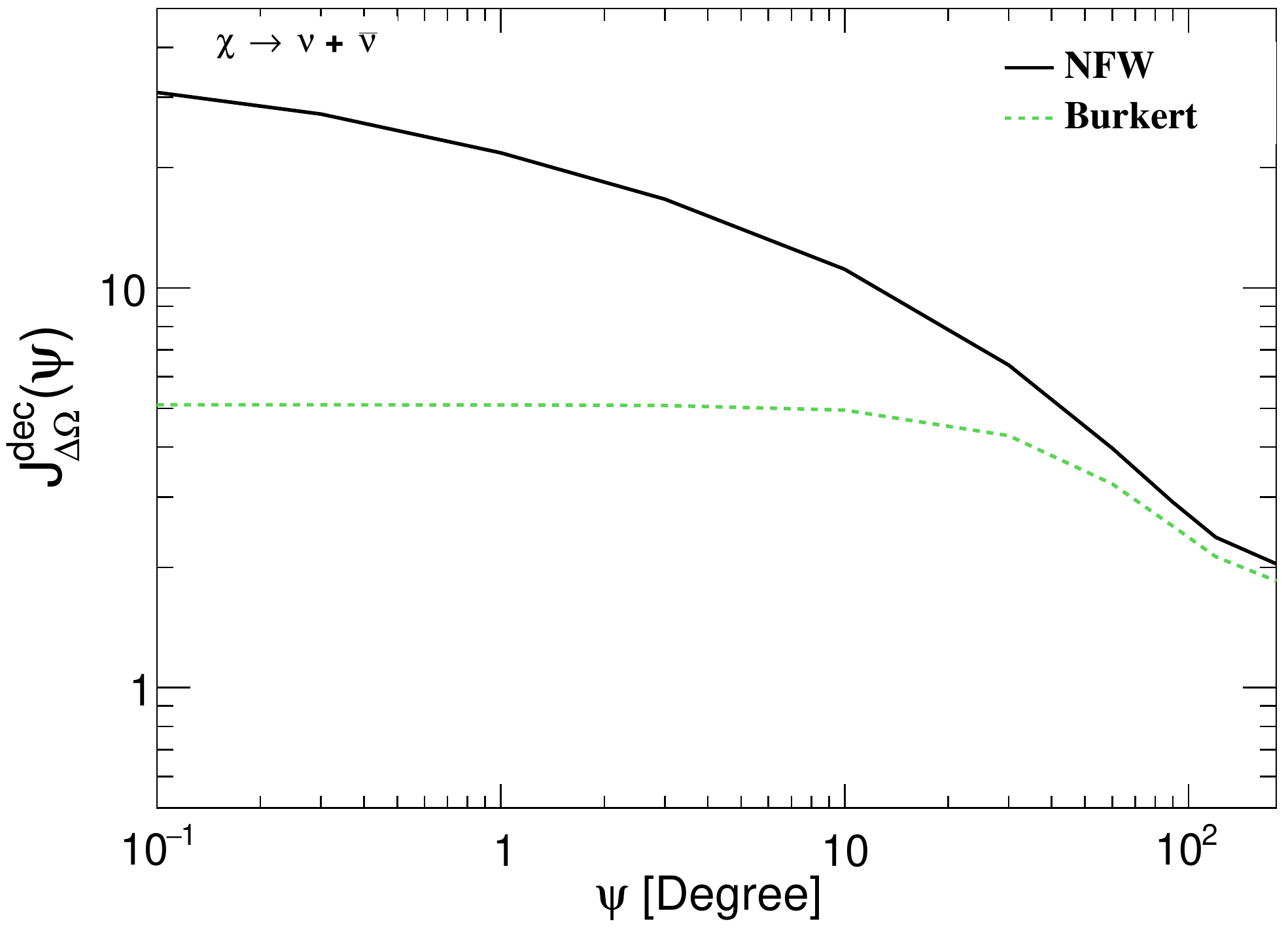}}
\mycaption{Line of sight integral for dark matter decay, $\mathcal{J}^{dec}(\psi)$, (see Eq.\,\ref{eq8}) vs. $\psi$ 
 and the average value of $\mathcal{J}^{dec}(\psi)$ over solid angle
$\Delta\Omega$, 
i.e., $\mathcal{J}_{\Delta\Omega}^{dec}(\psi)$ (see Eq.\,\ref{eq9}) for the decay process 
are shown in left and right panels respectively. In both the panels black solid and green dashed lines present the corresponding quantities 
for the NFW and the Burkert profiles respectively. We use the value of $\mathcal{J}^{dec}_{\Delta\Omega}$ ($\psi\,=\,180^\circ$) in our analysis, 
which are given by 2.04 and 1.85 for the NFW and Burkert profiles respectively.}
\label{fig3}
\end{figure}
For the decaying dark matter, $\mathcal{J}^{dec}(\psi)$ and $\mathcal{J}_{\Delta\Omega}^{dec}(\psi)$ are shown in left and 
right panels of Fig.\,\ref{fig3} respectively by the black solid (green dashed) lines using the NFW (Burkert) profile.
We obtain $\mathcal{J}^{dec}_{\Delta\Omega}$( $\psi\,=\,180^{\circ})$
= 2.04 and 1.85 for the NFW and Burkert profile respectively.  These agree with those presented in Ref.\,\cite{PalomaresRuiz:2007ry} up to uncertainties in the dark matter profile parameters.

The flux of neutrinos of each flavor per unit energy per unit solid angle in units of GeV$^{-1}$sr$^{-1}$cm$^{-2}$ s$^{-1}$ 
from the decay of dark matter particles is given by
\begin{equation}
 \frac{d^2\Phi^{dec}_{\nu/\bar\nu}}{{dE}\,d\Omega} = \mathcal{J}^{dec}_{\Delta\Omega}\,
 \frac{R_{sc}\rho_{sc}}{4\pi\,m_{\chi}\,\tau}\,\, \frac{1}{3} \frac{{dN^{dec}}}{{dE}}\,,
 \label{eq10}
\end{equation}
where $m_{\chi}$ is the mass of DM particle ($\chi$) in GeV, and $\tau$ is the decay lifetime of $\chi$
in second. The factor $\frac{1}{3}$ accounts for the averaging over total number of lepton flavors and 4$\pi$ 
implies isotropic decay. The mass of dark matter is   
shared by final $\nu$ and $\bar\nu$, thus, their energy spectrum can be written as
\begin{equation}
 \frac{{dN^{dec}}}{{dE}} = \delta(E_{\nu/\bar\nu} - m_\chi /2)\,.
 \label{eq11}
\end{equation}

\section{Key features of ICAL detector}
\label{sec:key features}
The proposed 50 kt MagICAL detector\,\cite{INO,Ahmed:2015jtv} is designed to have 151 alternate layers of 5.6 cm thick 
iron plates (act as target mass) and glass Resistive Plate Chambers (RPCs, act as active detector elements).
The plan is to have a modular structure for the detector with a dimension of 48\,m\,(L)\,$\times$\,16\,m\,(W)\,$\times$\,14.5\,m\,(H), 
subdivided into 3 modules, each having a dimension of 16\,m\,$\times$\,16\,m\,$\times$\,14.5\,m. The field strength of the magnetized
iron plates will be around 1.5 T, with fields greater than 1 T over at least 85$\%$ of the detector volume\,\cite{Behera:2014zca}.
Bending of charged particles in this magnetic field help us to identify the charges of $\mu^-$ and $\mu^+$ which are produced in the charged-current (CC) 
interactions of $\nu_\mu$ and $\bar\nu_\mu$ inside the detector. This magnetic field inside the detector is best suited to observe 
muons having energies in GeV range, measure their charges, and reconstruct their momentum with high precision\,\cite{Chatterjee:2014vta}.
The capabilities of ICAL to measure three flavor oscillation parameters based on the information coming from muon energy ($E_\mu$)
and direction ($\cos\theta_\mu$) have already been explored in Refs.\,\cite{Ghosh:2012px,Thakore:2013xqa}. Recently it has been
demonstrated that the ICAL detector has ability to detect hadron\footnote{These final state hadrons are produced along with the muons in 
CC deep-inelastic scattering processes in multi-GeV energies, and they carry vital information about the initial neutrino.} 
showers and extract information about hadron energy from 
them\,\cite{Devi:2013wxa,Mohan:2014qua}. The energy of hadron ($E^{/}_{had} = E_\nu - E_\mu$) can be calibrated using number of hits
in the detector due to hadron showers\,\cite{Devi:2013wxa}. In\,\cite{Devi:2014yaa}, it has been shown that by adding the hadron energy information 
($E^{/}_{had}$) to the muon information ($E_\mu$, $\cos\theta_\mu$) of each event the sensitivity of ICAL to the neutrino oscillation parameters
can be greatly enhanced. 
\begin{table}[htb!]
 \centering
\begin{tabular}{|l|l|}
\hline
Energy resolution ($\sigma_{E}$) (GeV) & $0.1\,\times$\,($E$/GeV)\\
\hline
Angular resolution ($\Delta \theta$) & $10^{\circ}$ \\
 \hline
Detection efficiency ($\mathcal{E}$) & $80\%$  \\
\hline
CID efficiency ($\mathcal{C}$) & $90\%$  \\
\hline
\end{tabular}
\mycaption{The detector characteristics used in the simulations. We use the same detector properties for $\mu^-$ and $\mu^+$ events.  }
\label{table1}
\end{table}
In this phenomenological study, we explore the physics reach of MagICAL to see the signatures of Galactic diffuse dark matter through neutrino portal 
using the neutrino energy ($E_\nu$) and zenith angle ($\cos\theta_\nu$) as reconstructed variables.
We consider reconstructed neutrino energy threshold to be 1 GeV for both $\mu^-$ and $\mu^+$ events. The energy resolution of the MagICAL 
detector is expected to be quite good, and we assume that the neutrino energy will be reconstructed with a Gaussian energy 
resolution of 10$\%$ of $E$/GeV (see table\,\ref{table1}). As far as the angular resolution is concerned, we use
a constant angular resolution of 10$^{\circ}$. For $\mu^{\mp}$ events, the constant detection efficiency is 80$\%$,
and the constant charge identification (CID) efficiency is 90$\%$. The detector properties that we use in our simulation 
agree quite well with the detector characteristics that have been considered in the existing phenomenological
studies related to the MagICAL detector. For example see Refs.\,\cite{Petcov:2005rv,Blennow:2012gj,Ghosh:2013zna, Ghosh:2014dba}.
We have checked that the representative choices of energy and angular resolutions of $\nu_\mu$ and $\bar\nu_{\mu}$ that 
we consider in this work can produce similar results 
for oscillation studies as obtained by the INO simulation code using muon momentum as variable.
In this work, we assume that the 50 kt MagICAL detector will collect atmospheric neutrino data for 10 years 
giving rise to a total exposure of 500 kt$\cdot$yr.

\section{Event spectrum and rates}
\label{sec:event spectrum}
In this section, we present the expected event spectra and total event rates at the MagICAL detector.
To estimate the number of expected $\mu^{-}$ events\footnote{The number of $\mu^{+}$ events from atmospheric neutrinos 
can be estimated using Eq.\,\ref{eq12} by considering appropriate flux, oscillation probability, cross-section, and detector properties.}
from atmospheric $\nu_{\mu}$ and $\bar\nu_{\mu}$\footnote{Atmospheric muon antineutrino flux gives rise 
to $\mu^{+}$ events in the detector, which can be misidentified as $\mu^{-}$ events.} in the $i$-th energy bin  
and $j$-th zenith bin at the MagICAL detector, we use the following expression\,\cite{Gandhi:2007td}

\begin{eqnarray}
N^{atm}_{ij}(\mu^-)\,=\,2\pi\,N_{\rm{t}} \, \mathcal{T} \int^{E^{i}_{max}}_{E^{i}_{min}} {dE^\prime}\, 
                  \int^{\cos\theta^{j}_{max}}_{\cos\theta^{j}_{min}}{d(\cos\theta^\prime)}\,
\int^{1}_{-1} d(\cos\theta) \int^\infty_{0} dE\,R({E,E^\prime})\nonumber\\\,R(\theta,\theta^\prime) \bigg[\,\sigma_{\nu_\mu}^{CC}({E})\, 
	   \mathcal{E}\,\mathcal{C}\,\bigg\{ \frac{{d^2}\Phi_{\nu_\mu} }{d\cos\theta\,d{E}}
	   P_{\mu\mu}\, +\, \frac{{d^2}\Phi_{\nu_e}}{d\cos\theta\,d{E}} P_{e\mu}\bigg\}\,+ \nonumber\\ \,\bar\sigma_{\nu_\mu}^{CC}({E})\, 
\mathcal{\bar E}\,(1-\mathcal{\bar C})\,\bigg\{ \frac{{d^2}\bar\Phi_{\nu_\mu} }{d\cos\theta\,d{E}}
\bar P_{\mu\mu}\, +\, \frac{{d^2}\bar\Phi_{\nu_e}}{d\cos\theta\,d{E}} \bar P_{e\mu}\bigg\}\bigg].\phantom{11}
  \label{eq12}
  \end{eqnarray} 
In the above equation, $\mathcal{T}$ is the total running time in second, and $N_t$ is the total number of target nucleons in the detector. 
The quantities $E$ ($E^\prime$) and $\theta$ ($\theta^\prime$) are the true (reconstructed) neutrino energy and zenith angle respectively.   
For $\mu^{-}$ ($\mu^{+}$) events, $\sigma_{\nu_\mu}^{CC}$ ($\bar\sigma_{\nu_\mu}^{CC}$) is the total neutrino (antineutrino) per nucleon 
CC cross-section. These cross-sections have been taken from Fig.\,9 of Ref.\,\cite{Formaggio:2013kya}.  
We take the unoscillated atmospheric $\nu_{\mu}$ and $\nu_{e}$ fluxes estimated for 
the INO site in units of m$^{-2}$s$^{-1}$GeV$^{-1}$ sr$^{-1}$ from Ref.\,\cite{Honda:2015fha,HONDA}.   
The probability of a $\nu_\mu$ ($\nu_{e}$) to survive (appear) as $\nu_\mu$ is denoted by $P_{\mu\mu}$ ($P_{e\mu}$).  
The  parameters $\mathcal{E}$ ($\mathcal{\bar E}$) and $\mathcal{C}$ ($\mathcal{\bar C}$) are the detection  
and charge identification efficiencies respectively for $\mu^-$ ($\mu^+$) events.  
The quantities $R(E,\,E^\prime) $ and $ R(\theta,\,\theta^\prime)$ are the Gaussian energy and angular resolution functions 
of the detector, which are expressed in the following way
\begin{equation}
 R(E,\,E^\prime) = \frac{1}{\sigma_{E}\sqrt{2\pi}}\,\,{\rm{exp}}\,\bigg\{\frac{-\,(E^\prime-E)^2}{2\sigma_{E}^2}\bigg\}
 \label{eq15},\,
\end{equation}
and
\begin{equation}
 R(\theta,\,\theta^\prime) = \frac{1}{\sigma_{\theta}\sqrt{2\pi}}\,\,{\rm{exp}}\,\bigg\{\frac{-\,(\cos\theta^\prime-\cos\theta)^2}{2\sigma_{\theta}^2}\bigg\}
 \label{eq16}\,.
\end{equation}
The parameters $\sigma_{E}$ and $\sigma_{\theta}\,(\sin\theta\,\Delta\theta$) denote the energy  
and angular resolutions as given in table\,\ref{table1}. 
\begin{table}[htb!]
 \centering
 \begin{tabular}{|l|l|l|l|}
  \hline
  Observables& Range & Width& Total bins \\
  \hline
  E$_{\nu}$ (GeV) & \makecell[c]{$1 , 15$\\$15 , 25$\\$25 , 50$ \\ $50 , 100$} & \makecell[c]{1 \\ 2 \\5 \\ 10} & $\left.\begin{tabular}{l}
14 \\ 5 \\ 5 \\ 5 \end{tabular}\right\}$  29\\
\hline
$\cos\theta_\nu$ & $-1, 1$ & 0.5 & 4\cr
\hline
 \end{tabular}
 \caption{The binning scheme adopted for the reconstructed $E_{\nu}$ and $\cos\theta_{\nu}$ for each muon polarity.
 The last column depicts the total number of bins considered for each observable.}
\label{tab:bin}
\end{table}

We can estimate the $\mu^-$ events in the $i$-th energy bin and $j$-th angular bin from the dark matter induced neutrinos and anitneutrinos
by making suitable changes in Eq.\,\ref{eq12} in the following fashion
\begin{eqnarray}
N^{dm}_{ij}(\mu^-)\,=\,2\pi\,N_{\rm{t}} \, \mathcal{T} \int^{E^{i}_{max}}_{E^{i}_{min}} {dE^\prime}\, 
                  \int^{\cos\theta^{j}_{max}}_{\cos\theta^{j}_{min}}{d(\cos\theta^\prime)}
\int^{1}_{-1} d(\cos\theta)\int^\infty_{0} dE\, R({E,E^\prime})\,\,R(\theta,\theta^\prime)\,\nonumber\\
           \frac{{d^2}\Phi^{dm} }{d\cos\theta\,d{E}} \bigg[\,\sigma_{\nu_\mu}^{CC}({E}) 
 	  \,\mathcal{E} \,\mathcal{C}\bigg\{P_{e\mu} + P_{\mu\mu} + P_{\tau\mu} \bigg\}\,+  \,\bar\sigma_{\nu_\mu}^{CC}({E}) 
  \,\mathcal{\bar E}\, (1-\mathcal{\bar C})\,\bigg\{\bar P_{e\mu} + \bar P_{\mu\mu} + \bar P_{\tau\mu} \bigg\}\bigg].\phantom{111}
  \label{eq13}
  \end{eqnarray} 
  
  In case of dark matter annihilation and decay, we have fluxes of $\nu_\tau$ and $\bar\nu_\tau$ along with the fluxes of 
  $\nu_e$, $\bar{\nu}_e$, $\nu_\mu$, and $\bar{\nu}_\mu$.  
The dark matter induced neutrino and antineutrino fluxes\footnote{The amount of  
$\nu_e$, $\bar{\nu}_e$, $\nu_\mu$, $\bar{\nu}_\mu$, $\nu_\tau$, and $\bar\nu_\tau$ fluxes from dark matter are same.} for each flavor  
are estimated  using Eqs.\,\ref{eq6} and \ref{eq10} for annihilation and decay processes respectively. 
In the above equation, the probability of $\nu_\tau$ ($\bar{\nu}_\tau$) to appear as $\nu_\mu$ ($\bar{\nu}_\mu$) at the detector is expressed by $P_{\tau\mu}$ ($\bar P_{\tau\mu}$). 
All the other symbols signify the same parameters as described in Eq.\,\ref{eq12}. In our analysis, we take $\delta_{\textrm{CP}}$ = 0$^\circ$ and therefore, 
we can write $P_{\alpha\beta}$ = $P_{\beta\alpha}$ and $\bar P_{\alpha\beta}$ = $\bar P_{\beta\alpha}$. 
Due to these properties and unitary nature of the PMNS matrix U\,\cite{Pontecorvo:1957qd, Maki:1962mu, Pontecorvo:1967fh}, 
the sums of oscillation probabilities for neutrino and antineutrino in above equation become 1. 
Therefore, $\nu_\mu$ and $\bar\nu_\mu$ event rates due to the dark matter annihilation/decay 
do not depend on the values of oscillation parameters. 

In our simulation, the full three flavor neutrino oscillation probabilities are incorporated using the PREM profile for the Earth matter density\,\cite{Dziewonski:1981xy}. 
The choices of central values of the oscillation parameters that are used in our simulation lie within the 1$\sigma$ range of these parameters
as obtained from the recent global fit studies\,\cite{Forero:2014bxa,Esteban:2016qun,Capozzi:2017ipn}.
We produce all the results in this paper using the following benchmark values of oscillation parameters: $\sin^2\theta_{23}$ = 0.5, $\sin^2 2\theta_{13}$=0.085, 
$\Delta m_{\textrm{eff}}^{2}$ = $\pm \, 2.4\,\times\,10^{-3}$ eV$^2$,  
$\sin^2 2\theta_{12}$ =  0.84, $\Delta m_{21}^{2}$ = 7.5\,$\times$\,10$^{-5}$ eV$^2$, and
 $\delta_{\textrm{CP}}$ = 0$^\circ$. The (+) and (-) signs of $\Delta m_{\textrm{eff}}^{2}$\footnote{The 
 effective mass-squared difference, $\Delta m_{\textrm{eff}}^{2}$, 
 is related to $\Delta m^{2}_{31}$ and $\Delta m^{2}_{21}$ through the expression\,\cite{Nunokawa:2005nx,deGouvea:2005hk}:
 \begin{eqnarray}
  \Delta m_{\textrm{eff}}^{2} = \Delta m^{2}_{31}\,-\Delta m^{2}_{21}(\cos^{2}\theta_{12}\,
				  -\,\cos\delta_{\rm CP}\, \sin\theta_{13}\,\sin 2\theta_{12}\,\tan\theta_{23})\,.			
 \end{eqnarray}} correspond to 
 normal ordering (NO) and inverted ordering (IO) respectively. 
 In fit, we keep the values of oscillation parameters and the choice of mass ordering fixed. 

In this analysis, we binned the $\nu$ and $\bar\nu$ data separately using reconstructed observables $E_{\nu}$ and $\cos\theta_{\nu}$ 
as described in table\,\ref{tab:bin}.
There are total 29 $E_{\nu}$ bins in the range of $E_{\nu}$\,=\,[1, 100] GeV. 
The bins of $E_{\nu}$ are chosen uneven to ensure that they are consistent with the energy resolution
of the detector at various energy ranges. The isotropic nature of the signal allows us to take coarser binning in $\cos\theta_{\nu}$, and 
we take four $\cos\theta_{\nu}$ bins of equal size in the range [-1, 1]. 
We use comparatively finer bins for reconstructed E$_{\nu}$ because the signal has a strong dependency on energy of neutrino. 
We adopt an optimized binning scheme so that we have at least 2 events in each bin. 
The total number of bins used in our analysis is 29\,$\times$\,4\,$=$\,116. We show the signal and background event distribution 
plots as a function of reconstructed neutrino energy for various $\cos\theta_\nu$ ranges in section\,\ref{sec:results} (see Figs.\,\ref{fig4} and \ref{fig6}).

\section{Simulation method}
\label{sec:simulation}
In our analysis, we consider the dark matter induced neutrinos as signal and treat atmospheric neutrinos as background.
If N$_{ij}^{atm}$ and N$_{ij}^{dm}$ denote the number of $\mu^{-}$ events produced from the interactions of atmospheric 
$\nu_\mu$ and dark matter induced $\nu_\mu$ respectively in the $i$-th energy and $j$-th angular bin (see Eqs.\,\ref{eq12} and \ref{eq13}), 
then the Poissonian $\chi^2$\,\cite{Olive:2016xmw} can be written as
\begin{eqnarray}
 \chi^2(\mu^{-}) =  \min_{\zeta_{atm},\,\zeta_{dm}}\sum^{N_{E_{\nu}}}_{i=1} \sum^{N_{\cos\theta_{\nu}}}_{j=1} 2\bigg[N^{th}_{ij}(\mu^{-}) - N_{ij}^{exp}(\mu^{-})\,- \,
 N_{ij}^{exp}(\mu^{-}) \ln \frac{N^{th}_{ij}(\mu^{-})}{N^{exp}_{ij}(\mu^{-})}\bigg]\nonumber\\+\,\zeta^2_{atm}\,+\,\zeta^2_{dm}
 \label{eq18}\,. \phantom{11}
\end{eqnarray}
In the above equation, $N_{ij}^{exp}\, =\, N^{atm}_{ij}$ and $N^{th}_{ij} = N^{atm}_{ij}\,(1 + \pi_{atm}\,\zeta_{atm}) + N^{dm}_{ij}\,(1 + \pi_{dm}\,\zeta_{dm})$ neglecting 
higher order terms. Here, $N_{E_{\nu}}$ = 29 and $N_{\cos\theta_{\nu}}$ = 4 as mentioned in table\,\ref{tab:bin}.  
The quantities $\pi_{dm}$ and $\pi_{atm}$ in Eq.\,\ref{eq18} are the over all normalization errors on signal and background respectively.
We take  $\pi_{dm}$ = $\pi_{atm}$\footnote{For a detailed discussion on the uncertainties of the atmospheric neutrino flux, see Ref.\,\cite{Honda:2006qj}.} = 20$\%$.
The systematic uncertainties in this analysis are 
incorporated using the pull method\,\cite{Huber:2002mx,Fogli:2002au,GonzalezGarcia:2004wg}. 
The parameters $\zeta_{dm}$ and $\zeta_{atm}$ are the pull variables due to the  
systematic uncertainties on signal and background respectively. 
The values of $\zeta_{dm}$ and $\zeta_{atm}$ are obtained by setting $\frac{\partial\chi^{2}}{\partial\zeta_{dm}}$\,=\,0 and 
$\frac{\partial\chi^{2}}{\partial\zeta_{atm}}$\,=\,0, and their values lie within the range -1 to 1. 
Following the same procedure, $\chi^2(\mu^+)$ for $\mu^+$ events is obtained. 
We calculate the total $\chi^2$ by adding the individual contributions from $\mu^-$ and $\mu^+$ events in the following way\footnote{Here, we would
like to mention that though we assume same amount of normalization 
uncertainties for $\mu^-$ and $\mu^+$ events, we get different values of $\zeta_{dm}$ and $\zeta_{atm}$ for $\mu^-$ and $\mu^+$ channels.}
\begin{equation}
 \chi^2_{\rm{total}} = \chi^2(\mu^{-})\, + \,\chi^2(\mu^{+})\,. 
 \label{eq19}
\end{equation}
We notice that our results remain unchanged if we consider larger uncertainties on the atmospheric neutrino events. The reason behind this is 
that for any choice of $m_\chi$ we have many bins in terms of the reconstructed observables E$_\nu$ and $\cos\theta_\nu$, which are not affected by the dark matter 
induced neutrinos. Therefore these bins can constrain the uncertainties on the atmospheric neutrino flux. 
On the other hand, we notice that if we take the larger uncertainties on the dark matter induced neutrino events, say 30$\%$,
our final results get modified by 2 to 3$\%$.  It is worthwhile to mention that the maximum uncertainty on the signal stems from the dark matter density profile. 
Therefore, we give our results assuming two different profiles for the dark matter density which are the NFW and the Burkert. 

As we have discussed in section\,\ref{sec:event spectrum}, the dark matter induced signal does not depend on the oscillation parameters
as long as we take the CP-violating phase $\delta_{\rm{CP}}$\,=\,0$^{\circ}$.
The dependency on the oscillation parameters in the results comes only through the atmospheric neutrino background. 
We produce all the results assuming normal ordering both in data and theory. We have checked that the results hardly change 
if we consider inverted ordering. One of the main reasons behind this is that due to our choice of coarser reconstructed $\cos\theta_\nu$ bins,
the information coming from the MSW effect\,\cite{Mikheev:1986gs,Mikheev:1986wj,Wolfenstein:1977ue,Wolfenstein:1979ni} in the 
atmospheric neutrino events gets smeared out substantially. Another reason is that since the dark matter 
induced neutrino signal appears only in 2 to 3 $E_\nu$ bins (see in Figs.\,\ref{fig4} and \ref{fig6}), 
$\chi^2$ is hardly affected due to the change in atmospheric neutrino background in these bins when we switch from NO to IO.

\section{Results}
\label{sec:results}
\subsection{Constraints on annihilation of dark Matter}
\label{subsec:resultA}
\begin{figure*}[htb!]
 \[
 \begin{array}{cc}
   \includegraphics[width=.49\linewidth]{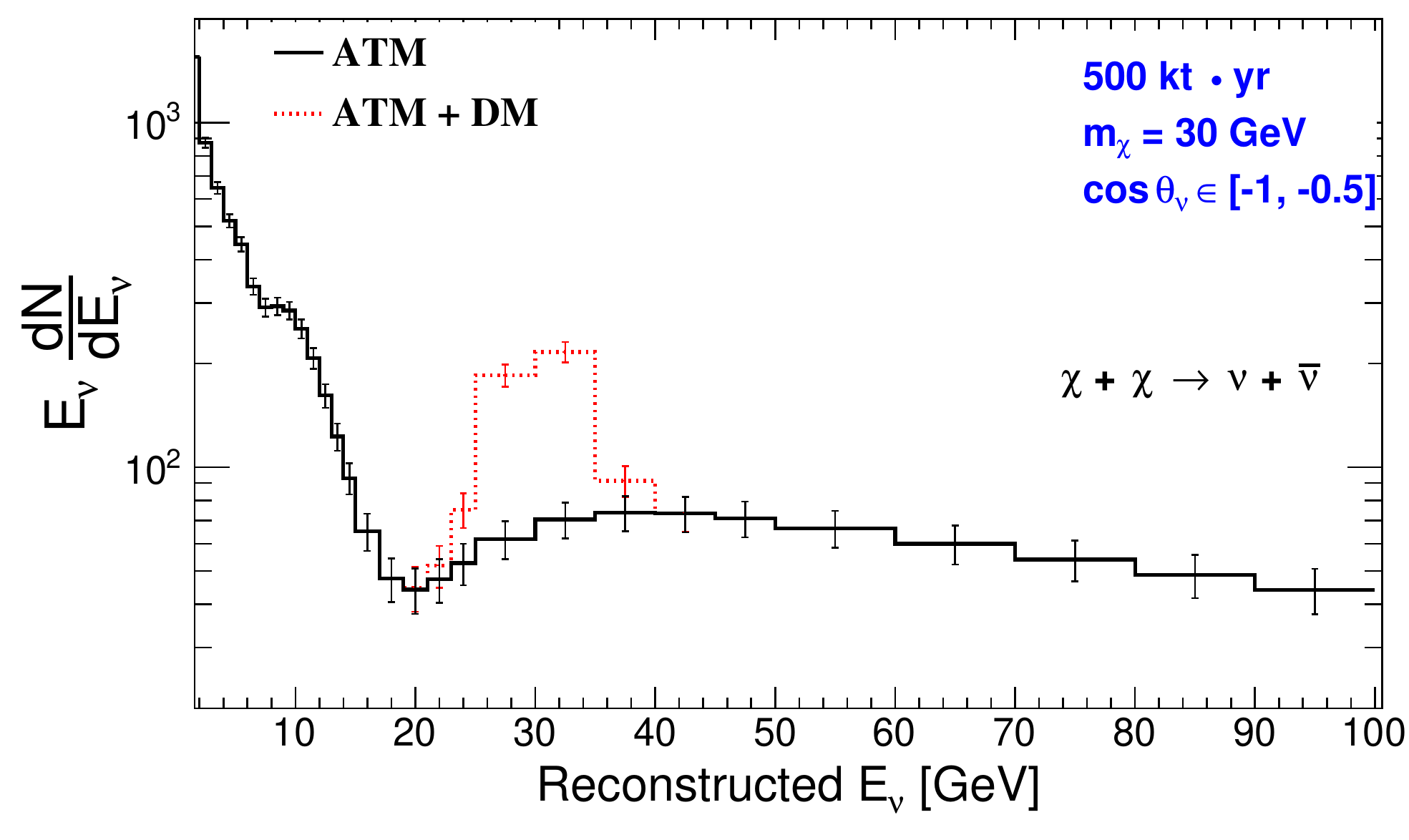}& 
   \includegraphics[width=.49\linewidth]{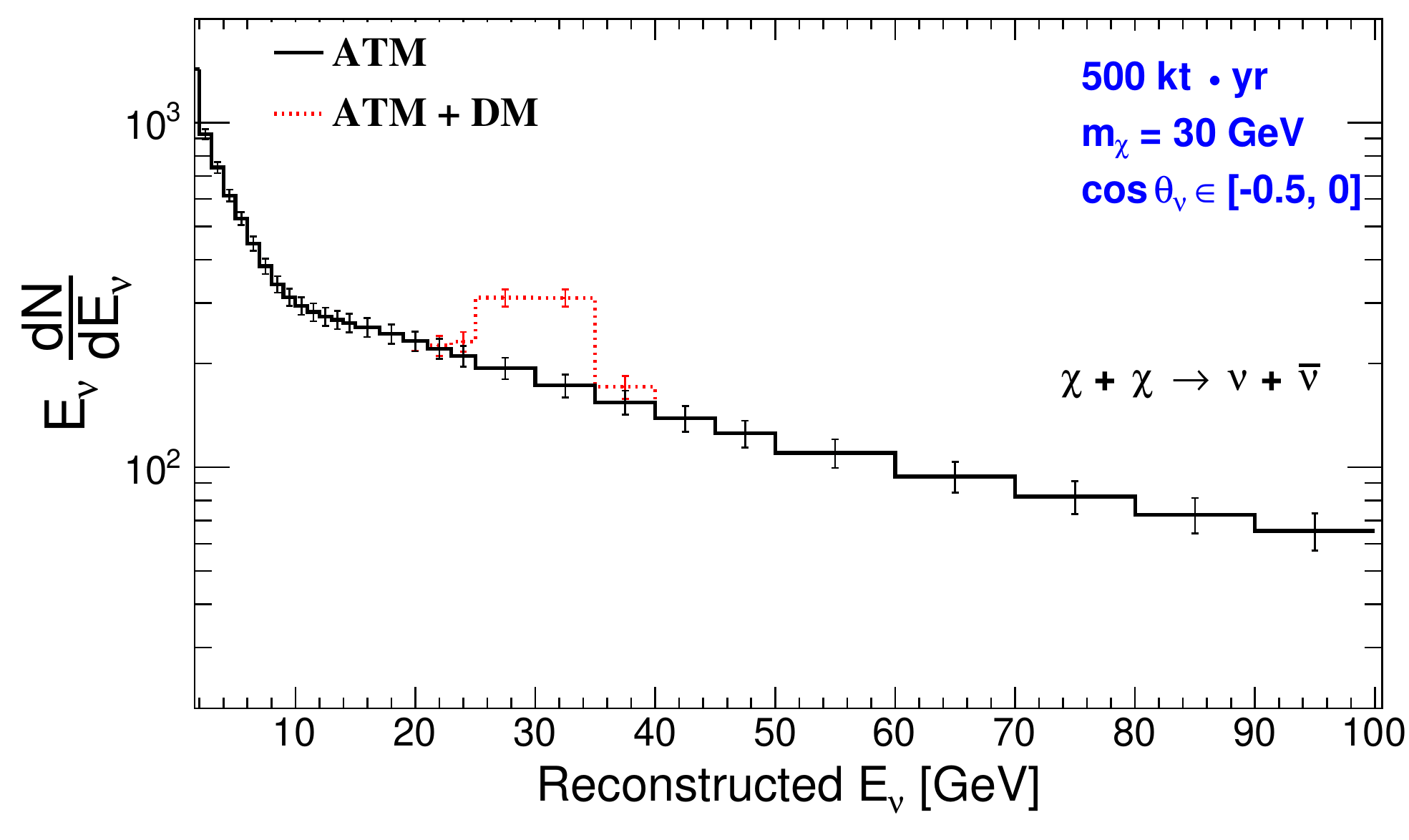}\\
   \includegraphics[width=.49\linewidth]{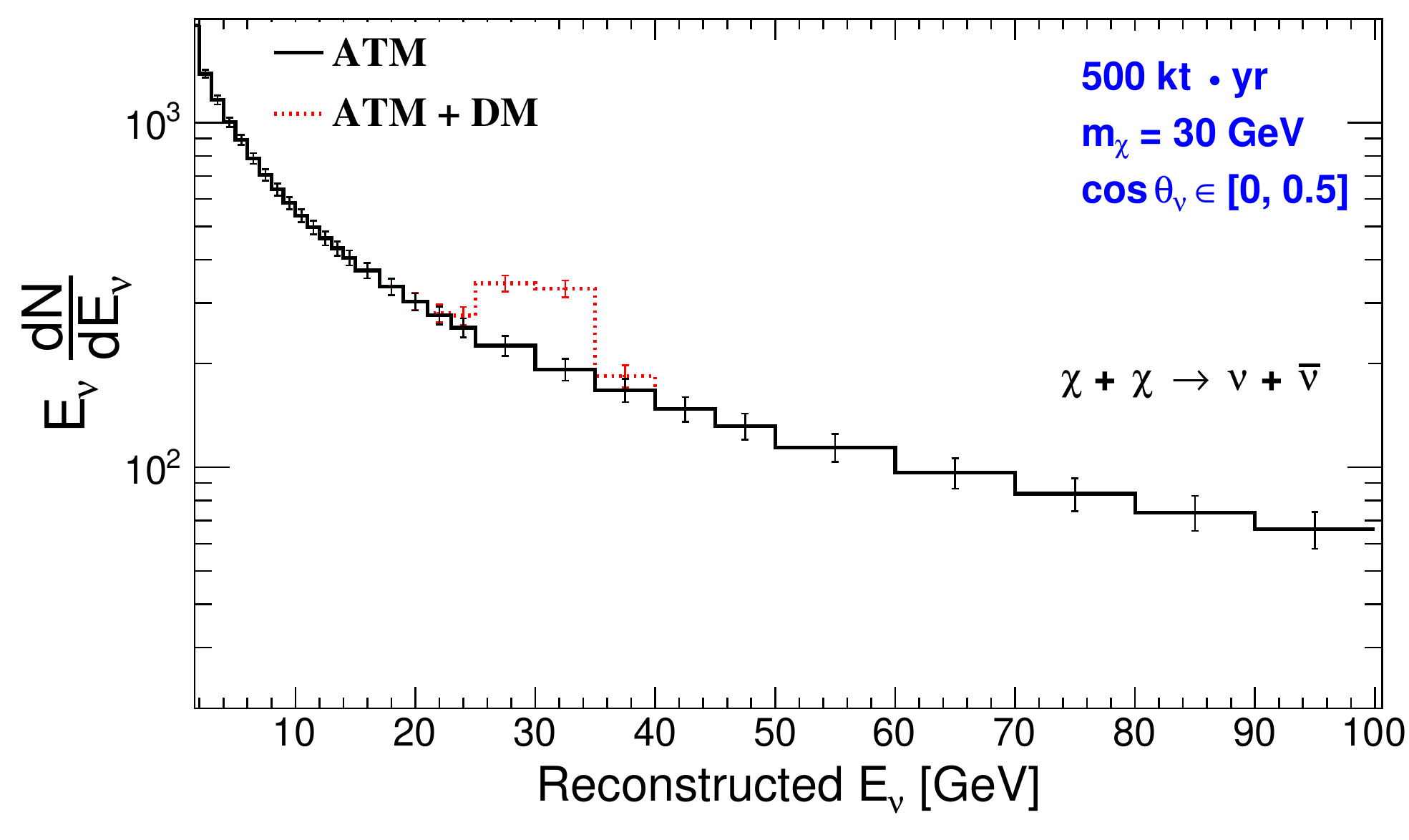}& 
   \includegraphics[width=.49\linewidth]{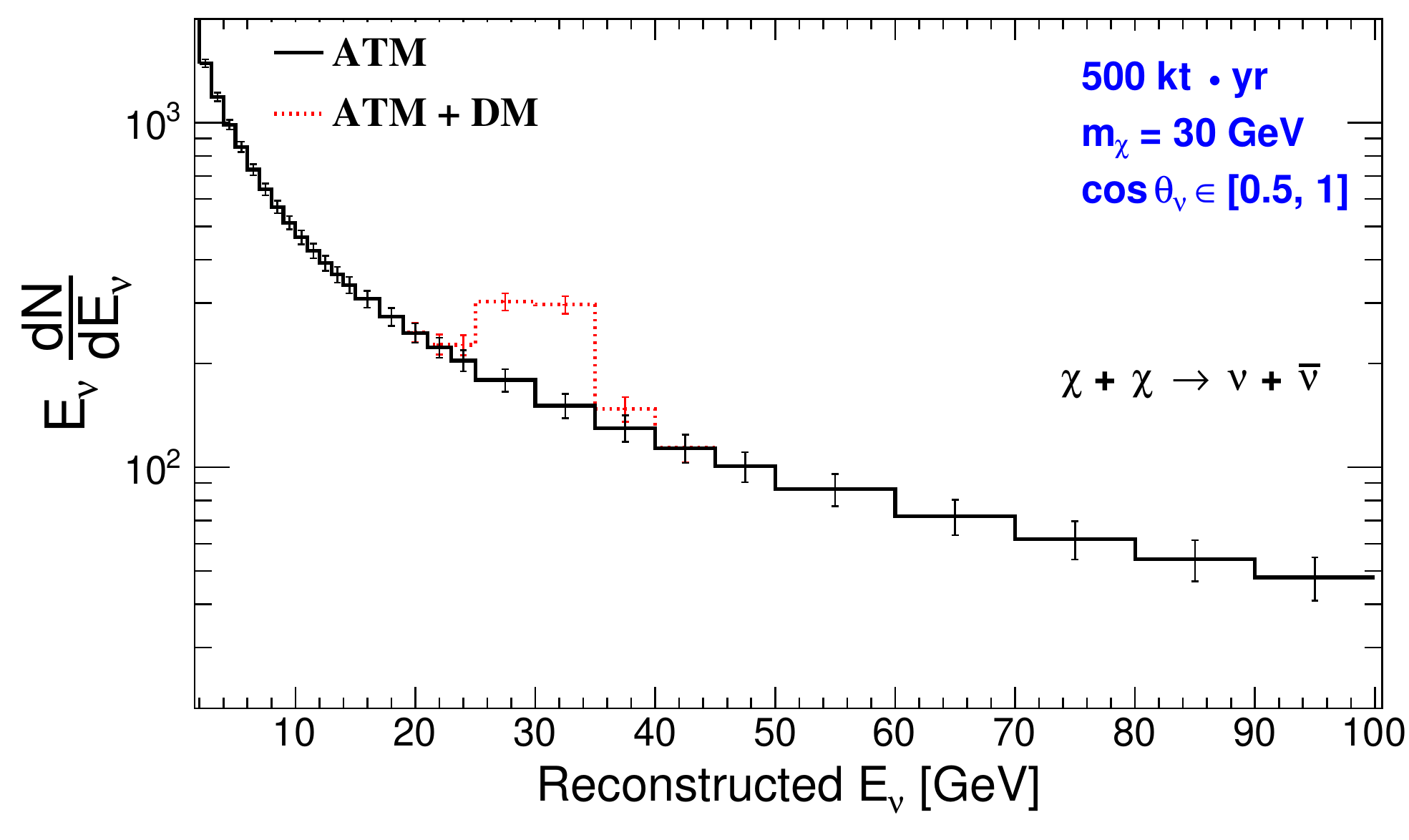}\\
   \end{array}
 \]
   \mycaption{Event spectra of atmospheric $\nu_\mu$ (denoted by ATM) are shown by the black
   solid lines. The predicted event distributions coming from atmospheric $\nu_\mu$ and dark
   matter originated neutrino (ATM + DM) are shown in red dotted lines for different 
   $\cos\theta$ ranges using 500 kt$\cdot$yr exposure of the MagICAL detector. The signal (DM)
   is coming from 30 GeV annihilating  DM particles here. The mass ordering is taken as NO. 
   The $\chi \chi \rightarrow \nu \bar{\nu}$ cross-section is arbitrarily chosen 
   to $<\sigma v> = 3.5\,\times\,10^{-23}$ cm$^3$\,s$^{-1}$ to have visual clarity. }
 \label{fig4}
  \end{figure*}

In this section, we present the constraints on self-annihilation cross-section of dark matter ($\chi \chi \rightarrow \nu \bar{\nu}$)
which can be obtained by 500 kt$\cdot$yr of MagICAL exposure. 
The background consists of conventional atmospheric neutrinos, and the signal consists of neutrinos coming from dark matter annihilation.
The simulated event spectra as a function of reconstructed neutrino energy in 500 kt$\cdot$yr exposure of MagICAL detector are presented 
in Fig.\,\ref{fig4}.  The quantity in the y-axis of Fig.\,\ref{fig4} is the number of
events per unit energy range multiplied by the mid value in each energy bin.    
In each panel, the black solid line represents the event distribution of conventional atmospheric $\nu_{\mu}$, denoted by ATM.
If DM particles of mass 30 GeV, for example, self-annihilate to $\nu\bar\nu$ pairs, 
then each of these $\nu$ and $\bar\nu$ will have 30 GeV of energy.  The  
total neutrino event spectra in MagICAL detector in presence of DM annihilation are shown 
by the red dotted lines (ATM + DM) in Fig.\,\ref{fig4}.  The value of self-annihilation cross-section of dark 
matter for these plots is taken to be 3.5 $\times \, 10^{-23}$ cm$^3$\,s$^{-1}$.

An excess of $\nu_\mu$ events due to dark matter annihilation appears over the ATM around reconstructed neutrino energy of 30 GeV.
These events get distributed over nearby energy bins due to the finite energy resolution of the 
detector. The number of signal and atmospheric events in neutrino mode are 174 and 210 
respectively in the energy range  [25, 35] GeV and $\cos\theta_\nu \in [-1, 1]$. 
There are 4 panels: each represents the event distribution summed over different $\cos\theta_\nu$ ranges.
The figures in top panels portray the event spectra over $\cos\theta_\nu \, \in \, [-1, -0.5]$ 
and [$-$0.5, 0.0] from left to right.  These events are due to upward going neutrinos, which travel a long 
distance through the Earth matter before they reach the detector. Though in these panels, the signatures 
of neutrino flavor oscillation are seen in ATM spectra, but the imprints of the Earth matter effect are not visible
due to the choice of our large $\cos\theta_\nu$ bins. 
The energy distributions of downward going events are shown in bottom panels from left to right for 
$\cos\theta_\nu$ $\in$ [0.0, 0.5] and [0.5, 1.0] respectively.  
These neutrinos do not oscillate as they traverse a length smaller than the oscillation wavelength 
in multi-GeV range.
The statistical error bars in these plots associated with different energy bins are equal to the square root of the number of events in the corresponding bins. 
  \begin{figure}[htb!]
\subfigure[]{\includegraphics[width =7.5cm]{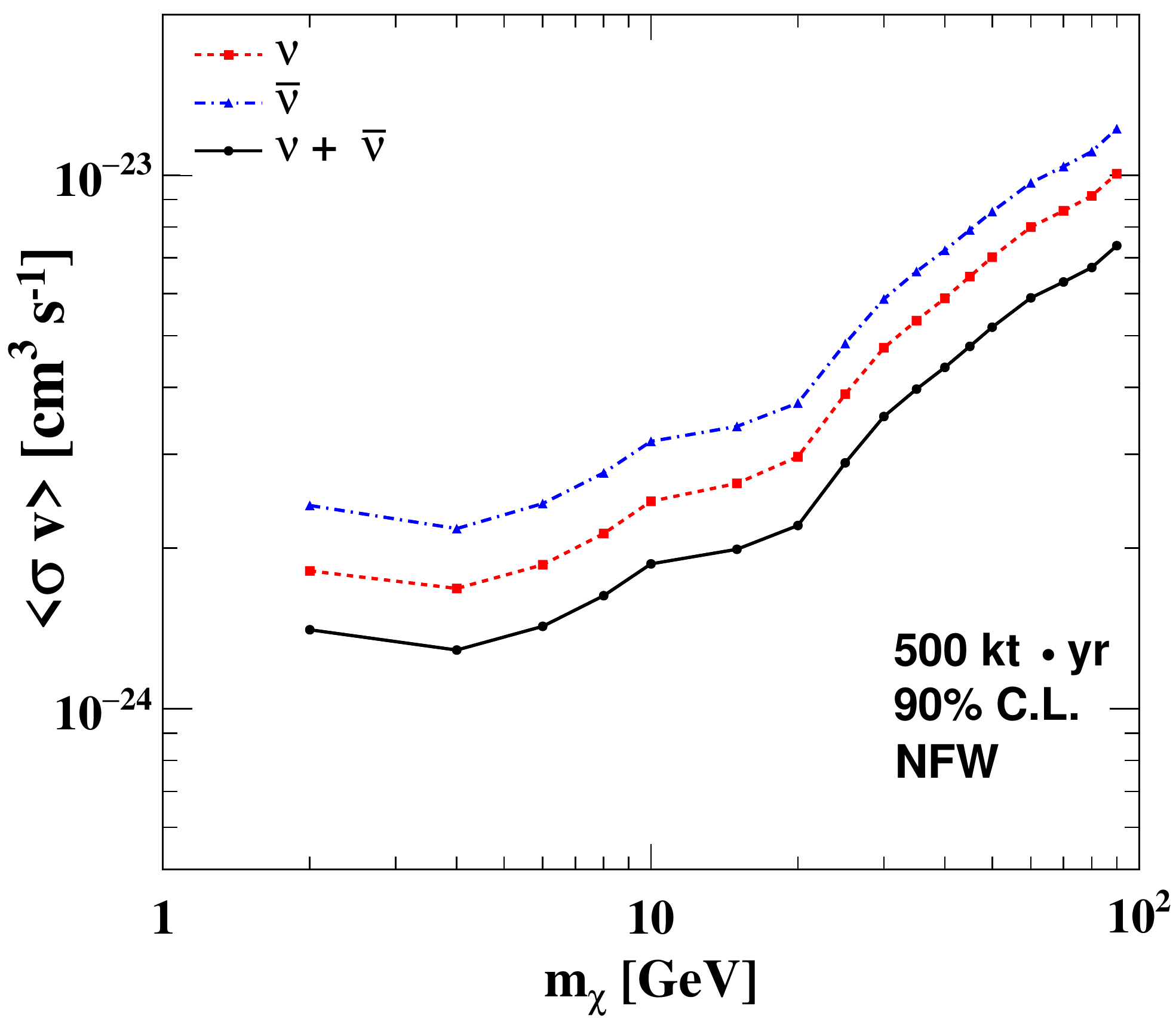}}
\subfigure[]{\includegraphics[width =7.5cm]{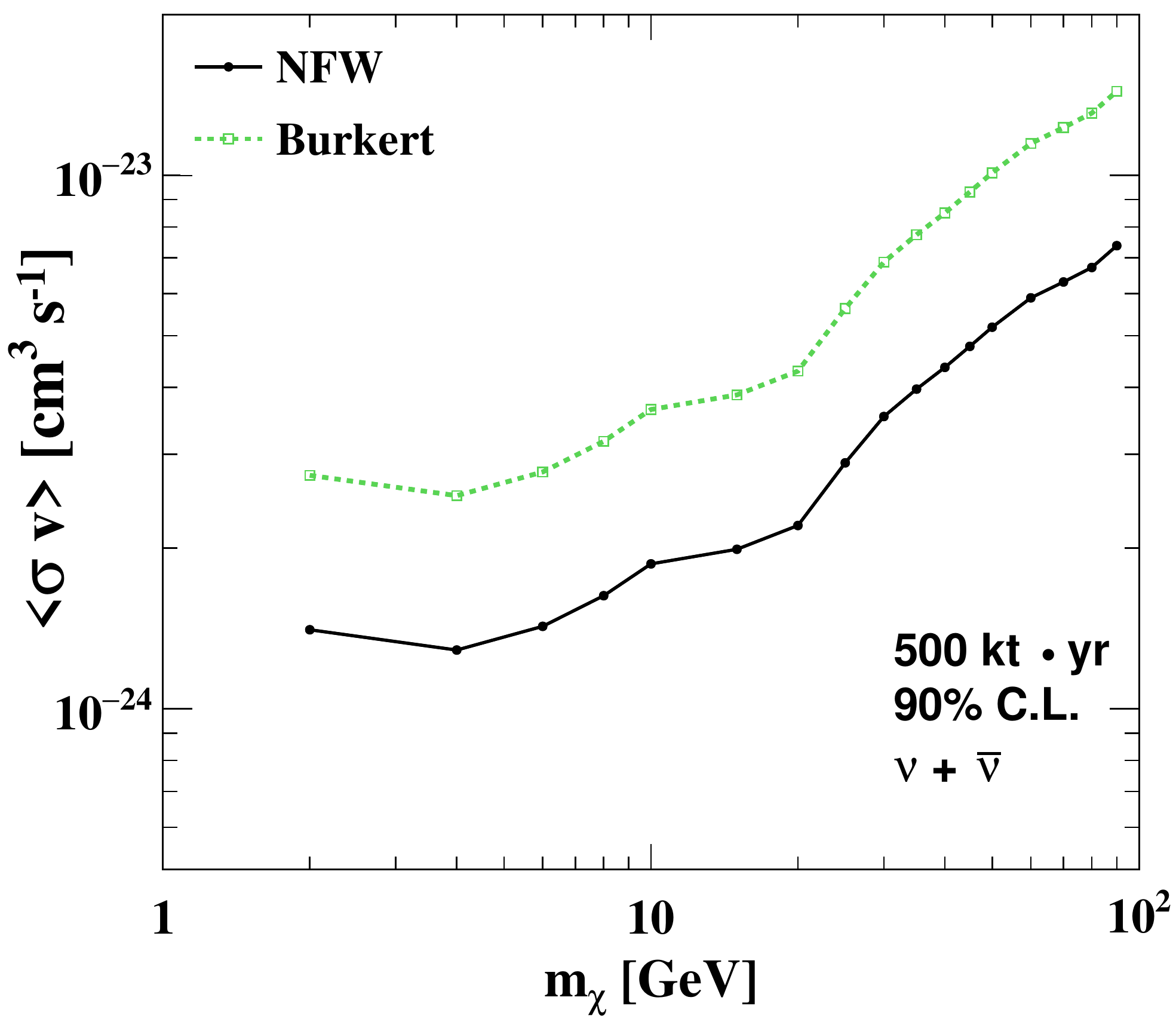}}
\mycaption{(a) The upper limit on self-annihilation cross-section of DM particle 
($\chi \chi \rightarrow \nu \bar{\nu}$) at 90$\%$ C.L. (1 d.o.f.) as a function of DM mass $m_\chi$ using 500 kt$\cdot$yr exposure of the MagICAL detector. 
The bound calculated with only $\nu_\mu$ ($\bar{\nu}_\mu$)-induced events is shown with red dashed (blue dot-dashed) line   
as MagICAL can distinguish $\nu_\mu$ from $\bar{\nu}_\mu$. The upper bound obtained by combining these two channels is also shown by the black solid line 
in the figure. We take the NFW as DM profile. (b) The upper 
bounds on the self-annihilation cross-section ($\chi \chi \rightarrow \nu \bar{\nu}$) of dark matter
are presented for the NFW (black solid) and Burkert (green dashed) profiles combining the information coming from $\nu_\mu$ and $\bar{\nu}_\mu$.
For both (a) and (b), the choice of mass ordering is NO.}
\label{fig5}
\end{figure}

The charge identification ability\footnote{We have checked that $\chi^2_\nu + \chi^2_{\bar\nu}$ is better than $\chi^2_{\nu + \bar\nu}$ by 
very little amount, which is around 2$\%$. In our analysis, CID does not play an important role 
unlike in the case of mass ordering determination for the following reasons. First, the signal is independent of oscillation parameters and it appears 
only in two to three $E_\nu$ bins. Secondly, the impact of the Earth matter effect in atmospheric 
$\nu$ and $\bar\nu$ events (background) gets reduced for our choices of large $\cos\theta_\nu$ bins.} of the MagICAL detector provides an opportunity to explore the 
same physics in neutrino and antineutrino channels separately.  This is not possible in water Cherenkov, liquid scintillator, and liquid argon
based detectors.  The MagICAL detector will have separate data sets for $\nu_\mu$ and $\bar{\nu}_\mu$. The total sensitivity is obtained
by combining the $\nu_\mu$ and $\bar{\nu}_\mu$ data sets according to Eq.\,\ref{eq19}.  We present results by using 
$\nu_\mu$ and $\bar{\nu}_\mu$ data separately, and then combining these two. The upper limits on self-annihilation cross-section 
($\langle\sigma v\rangle$) of DM particles for the process $\chi \chi \rightarrow \nu \bar{\nu}$ at 90$\%$ C.L. (1 d.o.f.) that 
MagICAL will obtain with  10 years of data are represented in Fig.\,\ref{fig5}.
The red dashed, blue dot-dashed, and the black solid lines in Fig.\,\ref{fig5}(a) represent
the limits on $\langle\sigma v\rangle$ from $\nu_\mu$, $\bar{\nu}_\mu$, and the combination of $\nu_\mu$ and $\bar{\nu}_\mu$ data respectively using the NFW profile. 
Analysis with $\nu_\mu$ gives tighter bound than $\bar{\nu}_\mu$ because
of the higher statistics of $\nu_\mu$ over $\bar{\nu}_\mu$. 

At higher energies, the atmospheric neutrino flux (background) decreases, and same happens
to the signal coming from dark matter self annihilation because of its $m^{-2}_{\chi}$ dependence (see Eq.\,\ref{eq6}). 
A competition between these two effects lowers the signal to background ratio for heavy dark matter particles. 
Thus, the bound on $\langle\sigma v\rangle$ becomes weaker for heavy DM. We can have a rough estimate of how 
 $\langle \sigma v\rangle$ depends on $m_\chi$ in the range say 4 to 8 GeV by mainly considering the energy dependence of flux and 
 interaction cross-section in both signal and background. In this $m_\chi$ range which also corresponds to neutrino energy range of 4 to 8 GeV,
 the atmospheric flux varies as $\sim E_\nu^{-2.7}$, whereas neutrinos flux from the annihilating DM goes as 
 $\langle \sigma v\rangle/m_\chi^{2}$. For both signal and background, the neutrino-nucleon CC cross-section is approximately 
 proportional to $E_\nu$ or $m_\chi$ in case of annihilation. Therefore, the neutrino signal from dark matter annihilation ($S$) depends on 
 $m_\chi$ in the following way: $S\,\propto \frac{\langle \sigma v\rangle}{m_\chi^{2}}\cdot m_\chi\,=\,\langle \sigma v\rangle/{m_\chi}$. 
 As far as background ($B$) is concerned, $B\propto m_\chi^{-2.7}\cdot m_\chi\,=\,m_\chi^{-1.7} $. Hence, in case of annihilation,  
 $S/\sqrt{ B} \propto \langle \sigma v\rangle \cdot m_\chi^{-0.15}$ or, $\langle \sigma v\rangle \propto m_\chi^{0.15}$ if $S/\sqrt{B}$ 
 remains constant. From Fig.\,\ref{fig5}(a), we can see that at $m_\chi$ = 4 GeV,  the limit on 
 $\langle \sigma v\rangle$ is 
 1.2\,$\times$\,10$^{-24}$ cm$^{3}$ s$^{-1}$ in case of  $\nu+\bar\nu$ modes. Now, from our approximate expression as mentioned above, the limit on $\langle \sigma v\rangle$ 
 at $m_\chi$ = 8 GeV should be around 1.2$\,\times\,$10$^{-24}\,\times\,(8/4)^{0.15}$ cm$^{3}$ s$^{-1}$ = $1.33\,\times\,10^{-24}$ cm$^3$ s$^{-1}$,
 which is indeed the case as can be seen from the black solid line in Fig.\,\ref{fig5}(a). If we want to do the same exercise
 for $m_\chi<$ 4 GeV, then the only change that we have to make is that the atmospheric neutrino flux varies as $E_\nu^{-2}$ at those energies 
 instead of $E_\nu^{-2.7}$. 
 On the other hand, to explain the nature of the same curve for $m_\chi$ above 8 GeV, we have to also take into account the effect of neutrino 
 flavor oscillation and detector response which have nontrivial dependence on $E_\nu$ whereas, the atmospheric neutrino flux still varies as $E_\nu^{-2.7}$
 in this energy range.

We compare the constraints with the NFW and the Burkert profiles by black solid and green dashed lines respectively in
Fig.\,\ref{fig5}(b) combining the neutrino and antineutrino data.  We obtain better sensitivity with the 
NFW profile than with the Burkert profile. The average value of $\mathcal{J}$ factor over 4$\pi$ solid 
angle for the Burkert profile is smaller than that for the NFW profile. Thus, the 
signal strength with Burkert profile is smaller than that with the NFW profile. We have  
$\mathcal{J}^{ann}_{\Delta\Omega}$ = 3.33 and 1.60 for the NFW and Burkert
profiles respectively, with $\Delta\Omega$ = 4$\pi$. 
\subsection{Constraints on decay of dark matter}
\label{subsec:resultB}

\begin{figure*}[htb!]
 \[
 \begin{array}{cc} 
   \includegraphics[angle=0.0,width=.49\textwidth]{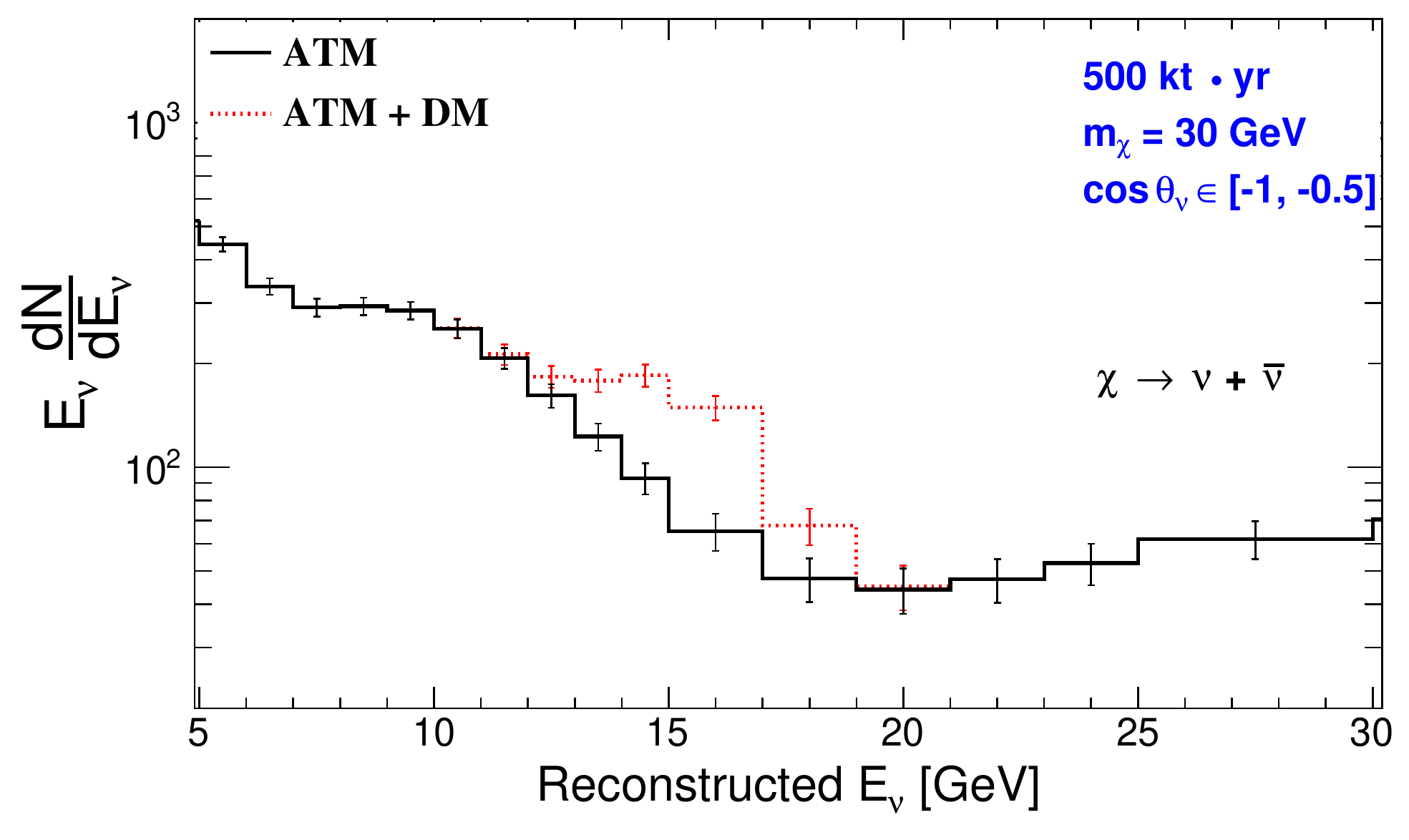}& 
   \includegraphics[angle=0.0,width=.49\textwidth]{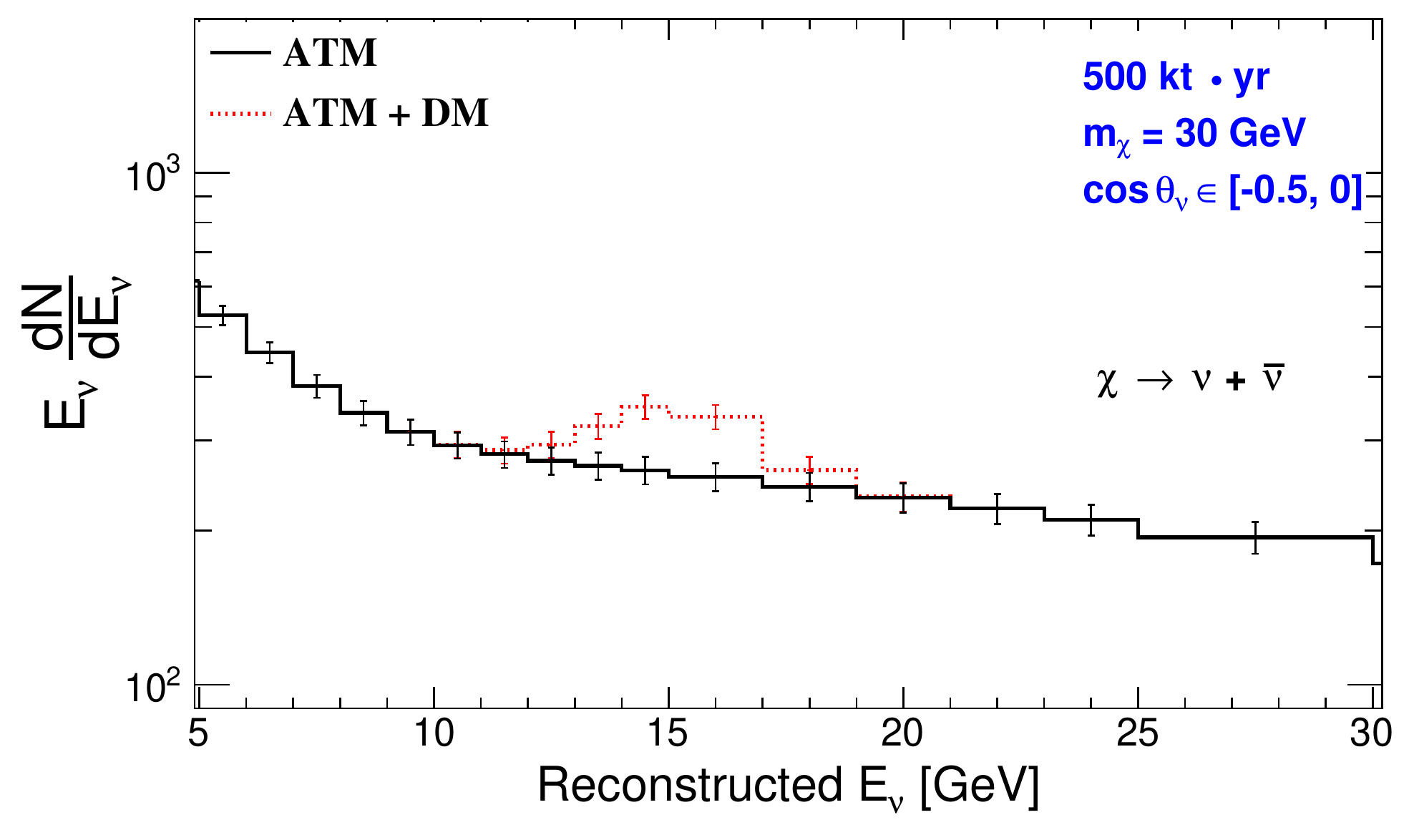}\\
   \includegraphics[angle=0.0,width=.49\textwidth]{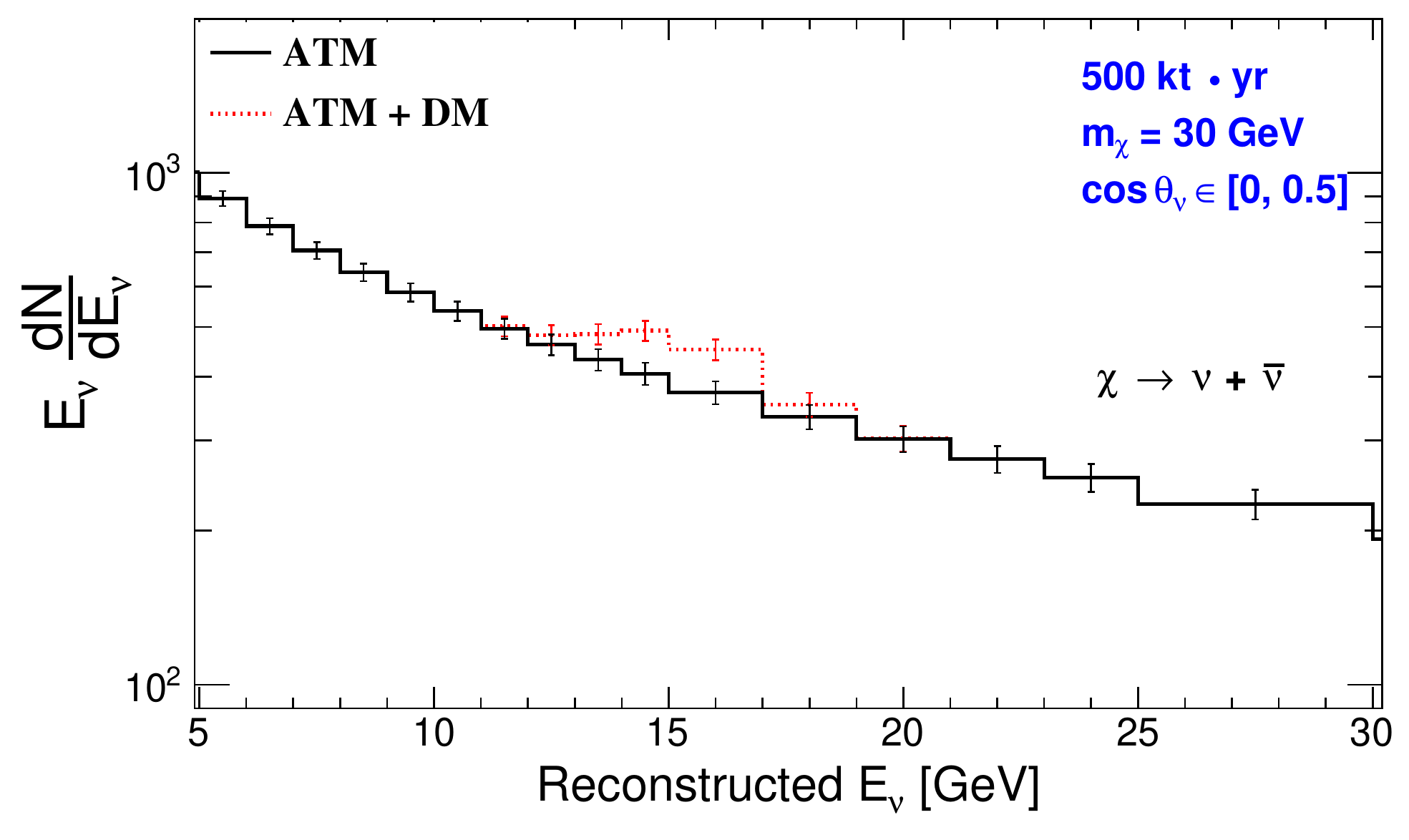}&
   \includegraphics[angle=0.0,width=.49\textwidth]{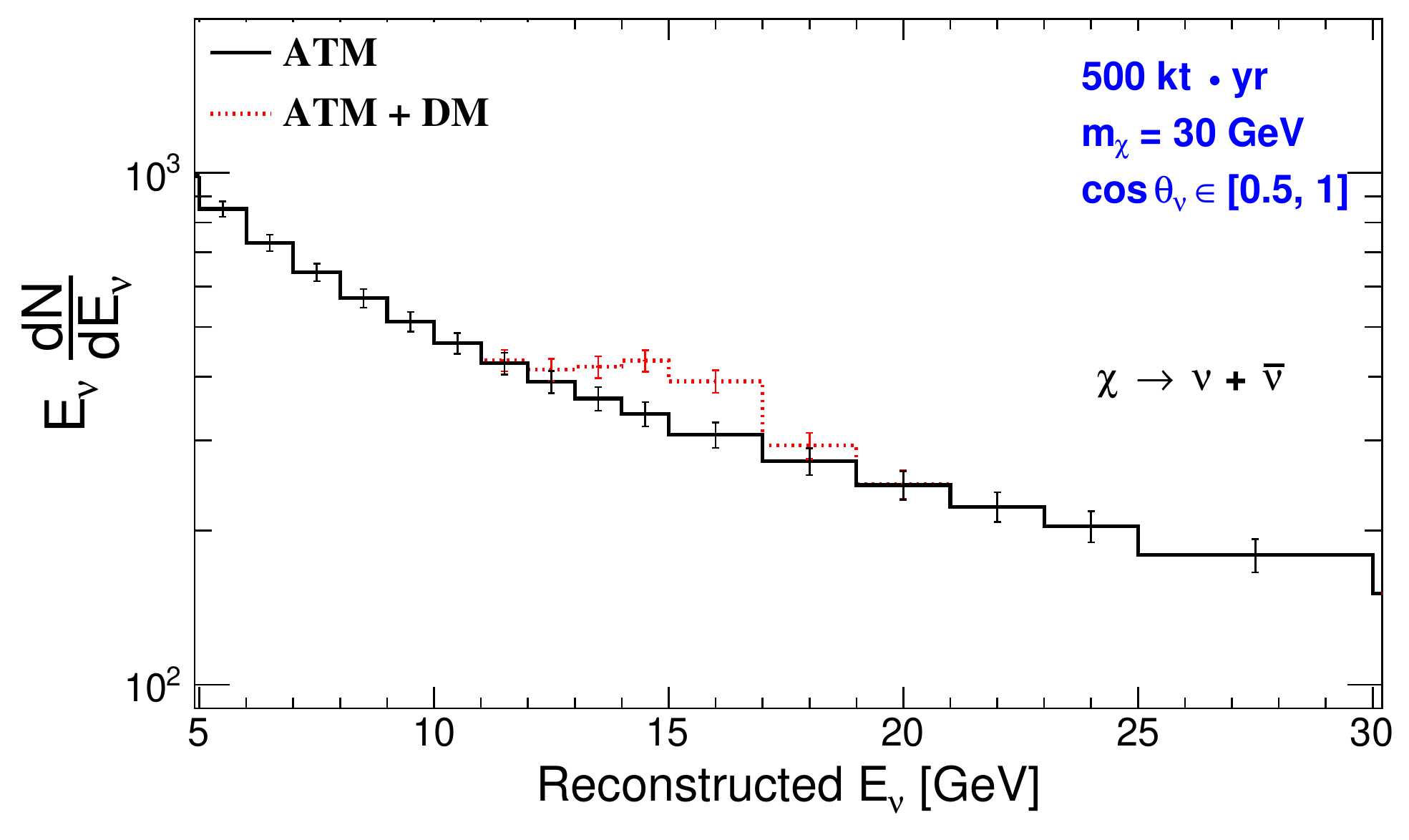}\\
 \end{array}
 \]
   \mycaption{The event distribution of atmospheric $\nu_\mu$ (denoted as ATM) and the predicted $\nu_\mu$ event spectra in presence of decay of 
  30 GeV dark matter particles (denoted as ATM + DM) in different $\cos\theta_\nu$ ranges using 500 kt$\cdot$yr exposure of the MagICAL detector. 
 Black solid (red dotted) line represents the ATM (ATM + DM). The mass ordering is taken as NO. 
 The lifetime of dark matter is arbitrarily chosen (4.7\,$\times\,10^{24}$ s) for sake of visual clarity. }
 \label{fig6}
 \end{figure*}

Assuming that dark matter particles have a mass of 30 GeV, and they decay to $\nu\bar\nu$ pairs, then the energy that each $\nu$ and $\bar\nu$ carries is 15 GeV. 
These events give rise to an excess of $\nu_{\mu}$ and $\bar\nu_\mu$ events around reconstructed neutrino energy of 15 GeV on top of the 
atmospheric neutrino event distribution as shown in Fig.\,\ref{fig6}. The black solid lines represent the event distributions
for the atmospheric neutrinos and the red dotted lines show event distributions for 
background along with the signal. The four panels in Fig.\,\ref{fig6} correspond to different $\cos\theta_\nu$ ranges as mentioned in the figure legends.
Here, we assume the lifetime ($\tau$) of dark matter to be 4.7\,$\times\,10^{24}$ s and we take 500 kt$\cdot$yr exposure for the MagICAL detector. 
We can see from Fig.\,\ref{fig6} that the events due to the decay of dark matter get distributed around 15 GeV due to the finite energy resolution of detector. 
In this case, the number of the signal and background events are 81 and 289 respectively in the reconstructed energy range [13, 17] GeV and $\cos\theta_\nu \in [-1, 1]$.
 \begin{figure}[htb!]
 \subfigure[]{\includegraphics[width =7.5cm]{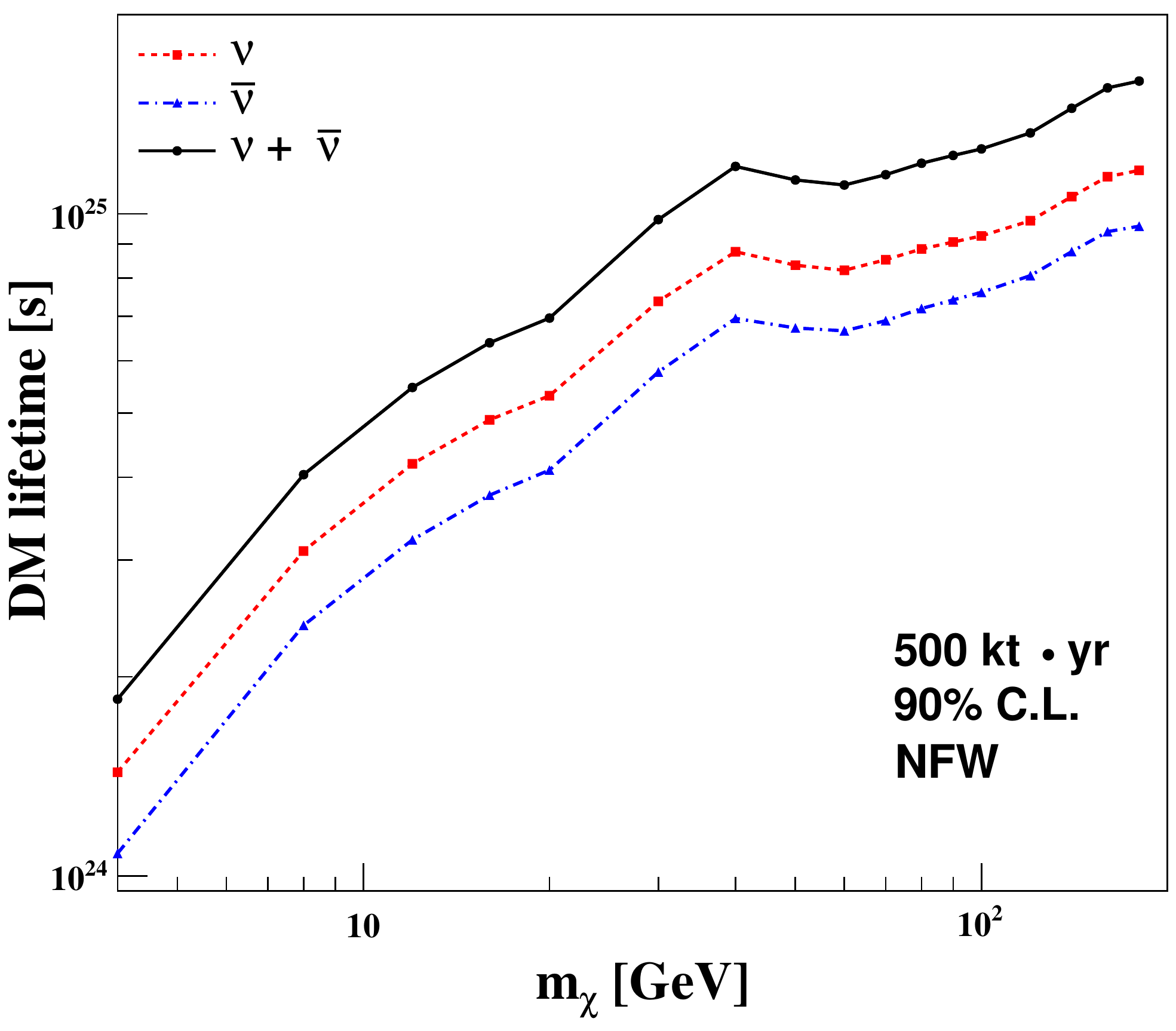}}
  \subfigure[]{\includegraphics[width =7.5cm]{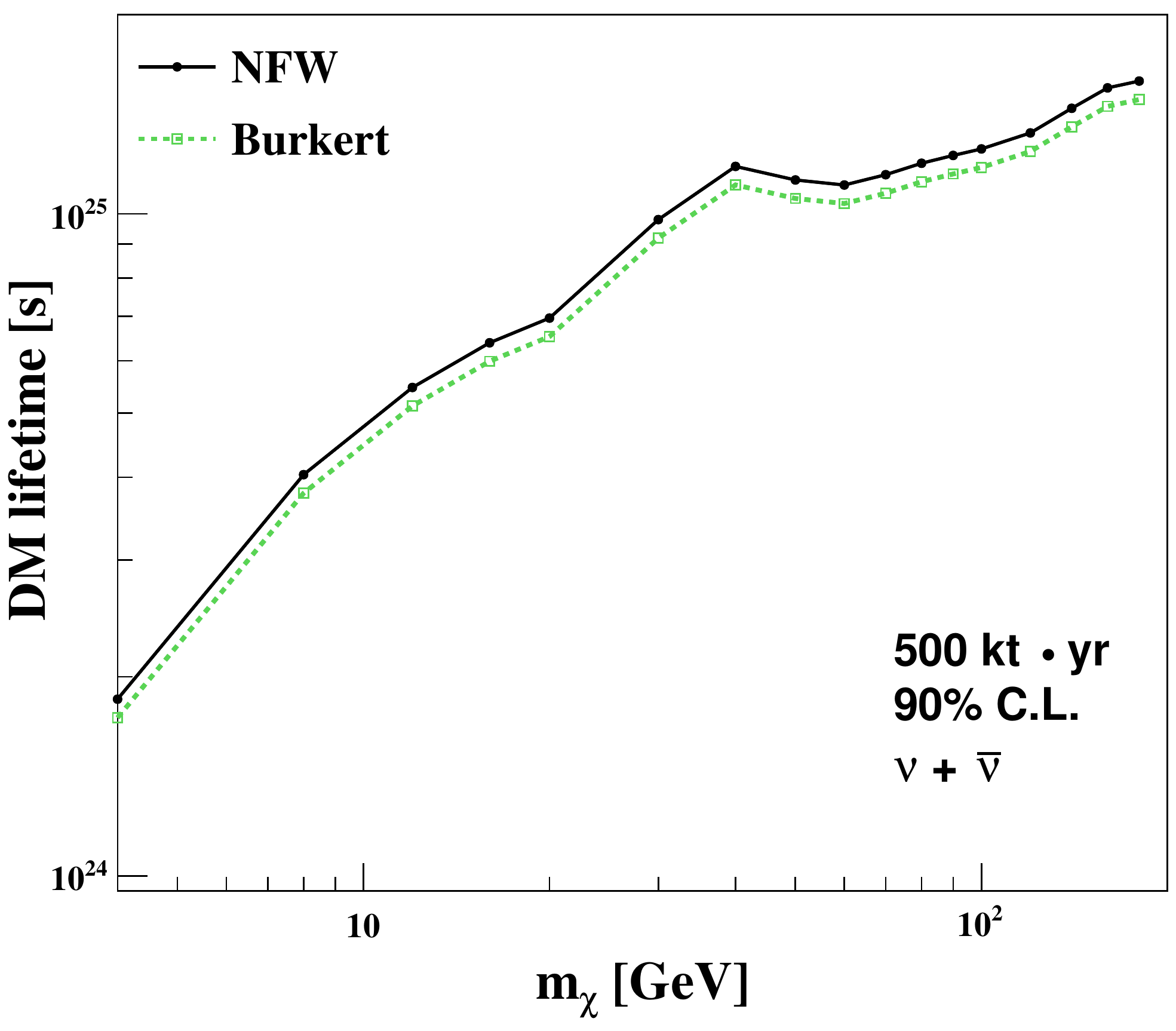}}
 \mycaption{(a) The lower bound on the decay lifetime of dark matter ($\chi \rightarrow \nu \bar{\nu}$)
 as a function of DM mass $m_\chi$ at 90$\%$ C.L. (1 d.o.f.) obtained using only $\nu_{\mu}$ 
 and only $\bar{\nu}_{\mu}$ data using 500 kt$\cdot$yr exposure of MagICAL. The red dashed (blue dot-dashed)
 line shows the  sensitivity coming from $\nu_\mu$ ($\bar{\nu}_\mu$)-induced events. The black solid line 
 represents the same combining $\nu_{\mu}$ and $\bar{\nu}_{\mu}$ events at $\chi^2$ level. We take NO as mass ordering.
 (b) The constraints on the decay lifetime of dark matter ($\chi \rightarrow \nu \bar{\nu}$)
 assuming the NFW (black solid line) and Burkert (green dashed line) profiles using 500 kt$\cdot$yr of MagICAL exposure.
 Here the results are shown combining $\nu$ and $\bar\nu$ (see section\,\ref{sec:simulation} for detail).}
\label{fig7}
\end{figure}

The future sensitivity of the MagICAL detector to set a lower limit on the lifetime ($\tau$) 
of dark matter as a function of $m_\chi$ is shown in Fig.\,\ref{fig7}. We give the results at 90$\%$ C.L. (1 d.o.f.) 
assuming 500 kt$\cdot$yr exposure of the proposed MagICAL detector. Here, we assume the dark matter density profile to be the NFW.
The red dashed (blue dot-dashed) line in Fig.\,\ref{fig7}(a) represents the bound which we obtain using $\nu_{\mu}$  ($\bar{\nu}_\mu$) data set.  
The bound gets improved when we add the $\nu_{\mu}$ and $\bar{\nu}_\mu$ data sets and the corresponding  
result is shown by the black solid line. Here, we see that the limits on the dark matter lifetime get improved when we consider higher m$_\chi$.
It happens for the following reasons. The flux of neutrinos coming from the dark matter decay (signal) has a $m_\chi^{-1}$ dependence (see Eq.\,\ref{eq10})
and the atmospheric neutrino flux (background) gets reduced substantially at higher energies. 
We find that in presence of these two competing effects, ultimately, the signal over background ratio gets 
improved for higher $m_\chi$, which allows us to place restrictive bounds on the lifetime of dark matter.
In Fig.\,\ref{fig7}(a), we can explain how the limit on dark matter life time $\tau$ depends on
$m_\chi$ in the range say 8 GeV $\leq$ $m\chi$ $\leq$ 16 GeV 
by mainly taking into account the energy dependence of the flux and neutrino-nucleon CC cross-section 
in the same fashion which adopt to explain the bound on $\langle\sigma v \rangle$ in the previous section.
The above range of $m_\chi$ corresponds to the $E_\nu$ range of 4 GeV to 8 GeV, since the neutrino energy from decaying DM is $E_\nu = m_\chi/2$.
Here, the neutrino flux from decaying DM is proportional to $\frac{1}{m_\chi\tau}$ (see Eq.\,\ref{eq10}). 
Thus, the neutrino signal ($S$) from dark matter decay varies as $S\propto \frac{1}{m_\chi\tau}\cdot m_\chi  = 1/\tau$, while the
background varies with $m_\chi$ in the same way as we see in case of 
annihilation which is $B\propto \,m_\chi^{-1.7}$. Hence, in case of decaying DM, $S/\sqrt{B}\propto \frac{1}{\tau}\cdot m_\chi^{0.85}$\,
or, $\tau\propto m_\chi^{0.85}$ for a fixed value of $S/\sqrt{B}$. From Fig\,\ref{fig7}(a), it can be seen that at
$m_\chi$ = 8 GeV, the limit on $\tau$ is 4.0$\,\times\,$10$^{24}$ s 
combining $\nu$ and $\bar\nu$ modes. From the simple $m_\chi$ dependence of $\tau$ that we discuss above, at 16 GeV, the limit on 
$\tau$ should be around  $4.0\,\times\,10^{24}\,\times\,(16/8)^{0.85}\,\textrm{s}\,=\,7.21\,\times\,10^{24}\,\textrm{s}$,
which is very close to the value as can be seen from the black solid line in  
Fig\,\ref{fig7}(a). To obtain the similar analytical understanding for $m_\chi$ below 8 GeV, we need to make suitable changes 
in the energy dependence of atmospheric neutrino flux which we have already discussed in the previous section.
Similarly, to see how $\tau$ varies with $m_\chi$ above 16 GeV, we have to also take into account the nontrivial
energy dependence of neutrino flavor conversion and detector response along with flux and cross-section.  

Due to the uncertainties in the dark matter density profiles, we present 
the bound on decay lifetime of dark matter with the profiles: NFW and Burkert by the black solid and 
green dashed line respectively in Fig.\,\ref{fig7}(b).  Ref.\,\cite{Dash:2014sza} considers only $\mu^+ \mu^-$ as final 
states for dark matter decay in the context of ICAL-INO, although their constraints are much weaker.
 
\subsection{Comparison with other experiments}
\begin{figure}[htb!]
\subfigure[]{ \includegraphics[width =7.5cm]{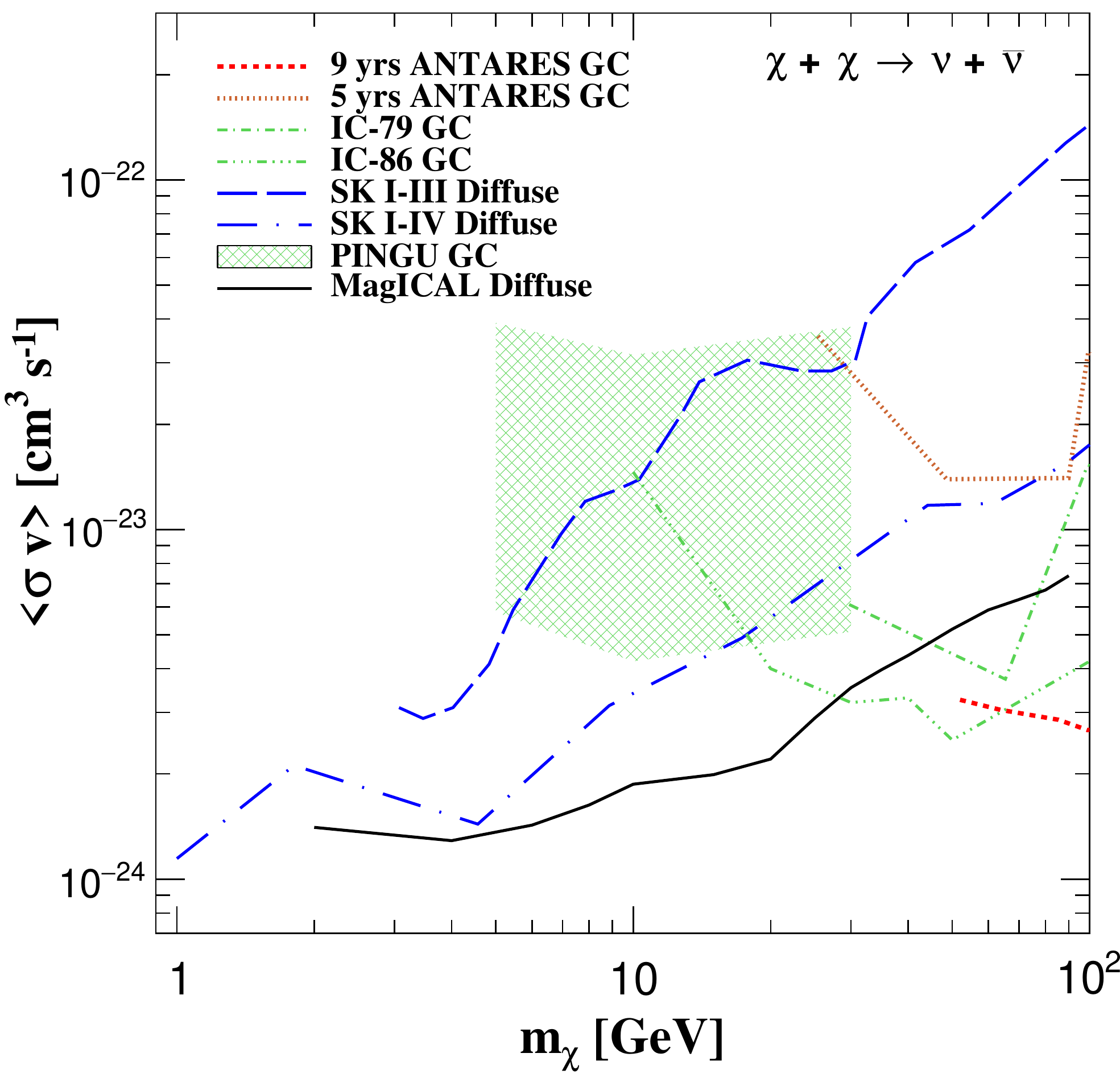}}
\subfigure[]{\includegraphics[width =7.5cm]{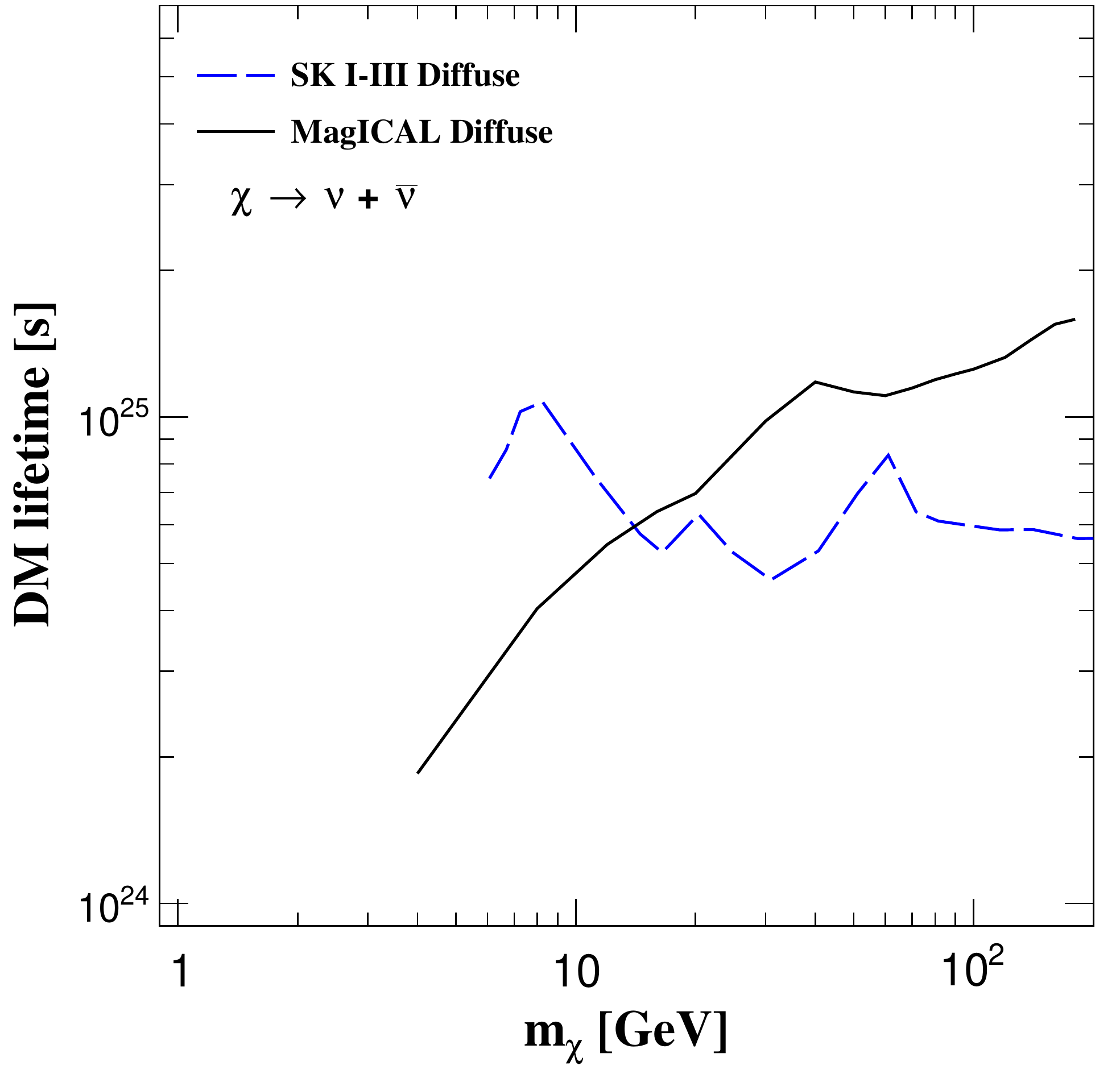}}
\mycaption{(a) Current bounds at 90$\%$ C.L. (1 d.o.f.) on self-annihilation cross-section which are obtained from the first three phases 
of Super-Kamiokande\,\cite{Mijakowski:2011zz} (blue long-dashed line), 
the four phases of Super-Kamiokande\,\cite{Mijakowski:2016cph} (blue long-dash-dotted line),
IceCube\,\cite{Aartsen:2015xej,Aartsen:2017ulx} (green dot-dashed and green triple-dot-dashed lines), 
and ANTARES\,\cite{Adrian-Martinez:2015wey,Albert:2016emp} (red dotted and red dashed lines) 
are shown. The future sensitivity of PINGU\,\cite{Aartsen:2014oha} with its 1 year of exposure is shown by green shaded region. 
We compare these limits with the bound obtained from 500 kt$\cdot$yr MagICAL (black solid line) detector.
For all the cases the NFW profile is used.
(b) Blue long-dashed line shows the current bound on decay lifetime of DM from the first three phases of Super-Kamiokande\,\cite{Mijakowski:2011zz} 
using the NFW profile. We compare this limit with the performance of 500 kt$\cdot$yr MagICAL detector (black solid line) using 
the same NFW profile.}
\label{fig8}
\end{figure}
Various experiments present the bounds on the self-annihilation cross-section of $\chi\chi\rightarrow\nu\bar\nu$ 
and the decay lifetime of $\chi\rightarrow\nu\bar\nu$ processes. Fig.\,\ref{fig8}(a) shows 
a comparison of the current bounds on $\langle \sigma v \rangle$ at 90$\%$ C.L. (1 d.o.f.)
from the first three phases of the Super-Kamiokande\,\cite{Mijakowski:2011zz} (blue long-dashed line),  
the four phases of the Super-Kamiokande\,\cite{Mijakowski:2016cph} (blue long-dash-dotted line), 
IceCube\,\cite{Aartsen:2015xej,Aartsen:2017ulx} (green dot-dashed and green triple-dot-dashed lines),
ANTARES\,\cite{Adrian-Martinez:2015wey,Albert:2016emp} 
(red dotted and red dashed lines), 
PINGU\,\cite{Aartsen:2014oha} (green shade), and from the MagICAL detector (black solid line) for the process $\chi\chi\rightarrow\nu\bar\nu$. 
We do not show the weaker limits from Baikal NT200\,\cite{Avrorin:2016yhw}. 
In Fig.\,\ref{fig8}(b), we compare the limits on decay lifetime ($\tau$) for the process $\chi \rightarrow \nu \bar{\nu}$ from 
the first three phases of the Super-Kamiokande experiment\,\cite{Mijakowski:2011zz} (blue long-dashed line) 
and the present work (black solid line).

Due to the lower energy threshold of MagICAL, the dark matter constraints can be estimated for $m_\chi$ values which are as low 
as 2 GeV and 4 GeV in case of annihilating and decaying dark matter respectively. The good energy and direction resolutions
of MagICAL detector help to strongly constrain the $\langle \sigma v\rangle$ and $\tau$ for $m_\chi$ in multi-GeV range. 
The constraints on $ \langle\sigma v\rangle$ obtained using 
319.7 live-days of data from IceCube operating in its 79 string configuration during 2010 and 2011 are stronger than MagICAL for 
dark matter masses heavier than $\sim$ 50 GeV (see green dot-dashed line in Fig.\,\ref{fig8}(a))\,\cite{Abbasi:2011eq,Dasgupta:2012bd,
 Aartsen:2013dxa,Aartsen:2014hva,Moline:2014xua,Rott:2014kfa,Aisati:2015vma,Aartsen:2015xej,Chianese:2016opp,Boucenna:2015tra,
 Aisati:2015ova,Aartsen:2016pfc}. But, if we consider the limits on $\langle \sigma v\rangle$ estimated using 
 three years of the IceCube/DeepCore data\,\cite{Aartsen:2017ulx}, then their performance becomes better than the MagICAL 
 detector for $m_\chi\geq 30$ GeV (see green triple-dot-dashed line in Fig.\,\ref{fig8}(a)).  
Using the 9 years data of ANTARES, no excess was found over the expected neutrino events 
in the range of WIMP mass 50 GeV\,$\leq m_\chi\leq$\,100 GeV, and they presented the most stringent constraint on 
$\langle \sigma v \rangle$ for $m_\chi\geq$ 70 GeV\,\cite{Albert:2016emp}.   However, for dark matter masses $\lesssim$ 100 GeV, the potential 
constraints from MagICAL are comparable or slightly better than that from Super-Kamiokande\,\cite{Mijakowski:2011zz,Mijakowski:2016cph}. 
The limit on $\langle \sigma v\rangle$ by 500 kt$\cdot$yr exposure of MagICAL detector is better than that from 1 year exposure of PINGU\,\cite{Aartsen:2014oha}. 
The constraints on dark matter annihilation and decay that we show in Fig.\,\ref{fig8} can only be obtained from neutrino telescopes, including liquid 
scintillator detectors\,\cite{Kumar:2015nja,Wurm:2011zn}.  The dark matter masses that we consider are too low for efficient electroweak 
bremsstrahlung, and hence gamma-ray constraints on this channel are weak\,\cite{Kachelriess:2007aj,Bell:2008ey,
Bell:2011eu,Bell:2011if,Cirelli:2010xx,Murase:2015gea,Esmaili:2015xpa,Chowdhury:2016bxs,Queiroz:2016zwd}.  Since MagICAL can distinguish 
between $\mu^+$ and $\mu^-$, it can also give constraints on exotic lepton number violating dark matter 
interactions.  The potential dark matter constraints from Baikal-GVD, and Hyper-Kamiokande will be stronger or comparable\,\cite{Avrorin:2014vca,Abe:2011ts}.  The 
complementarity of INO-MagICAL with PINGU and Hyper-Kamiokande will certainly make dark matter physics richer.

\subsection{The constraints on DM-induced neutrino flux}
\begin{figure}[htb!]
\subfigure[]{\includegraphics[width=7.5cm]{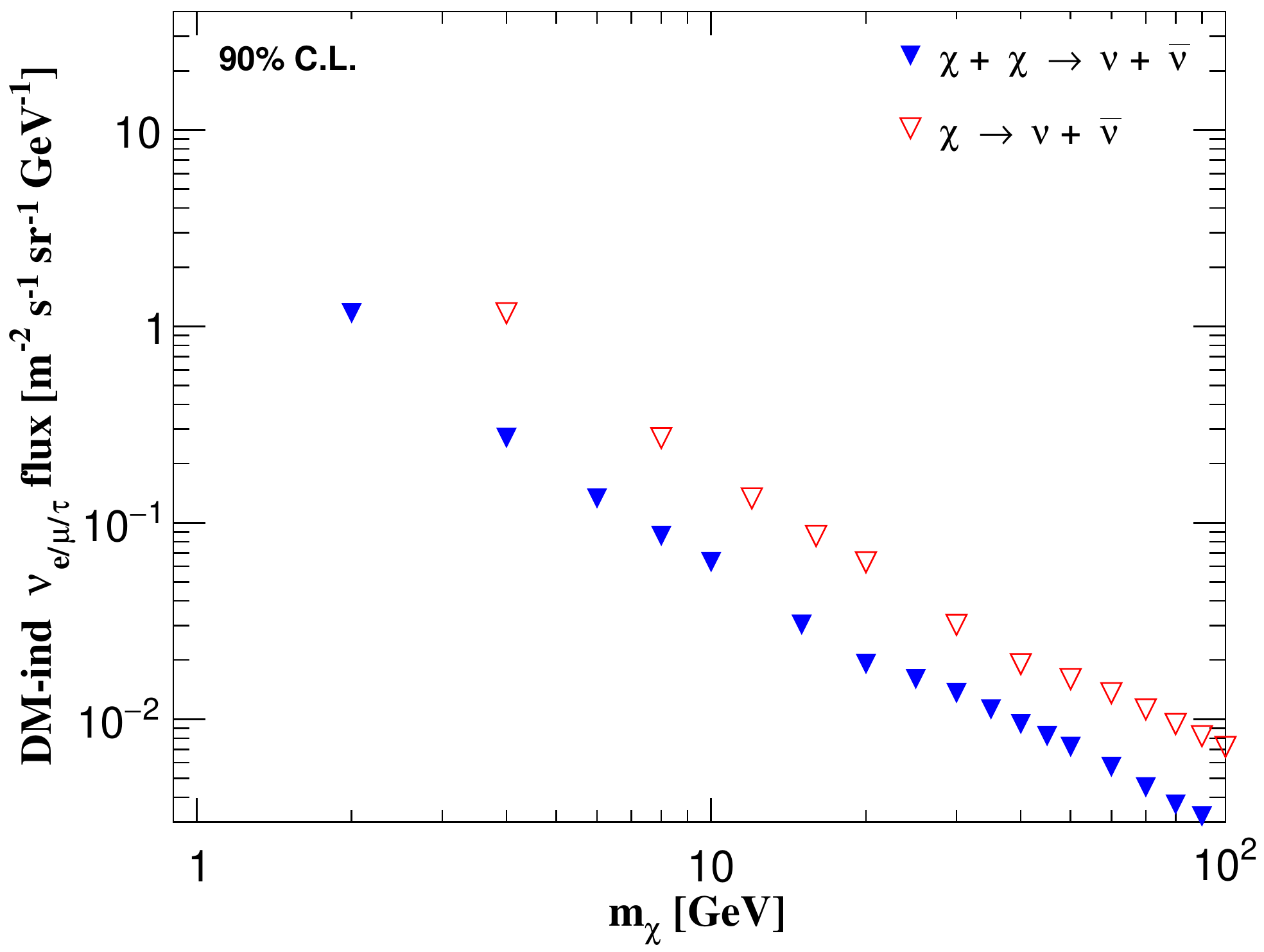}}
\subfigure[]{\includegraphics[width=7.5cm]{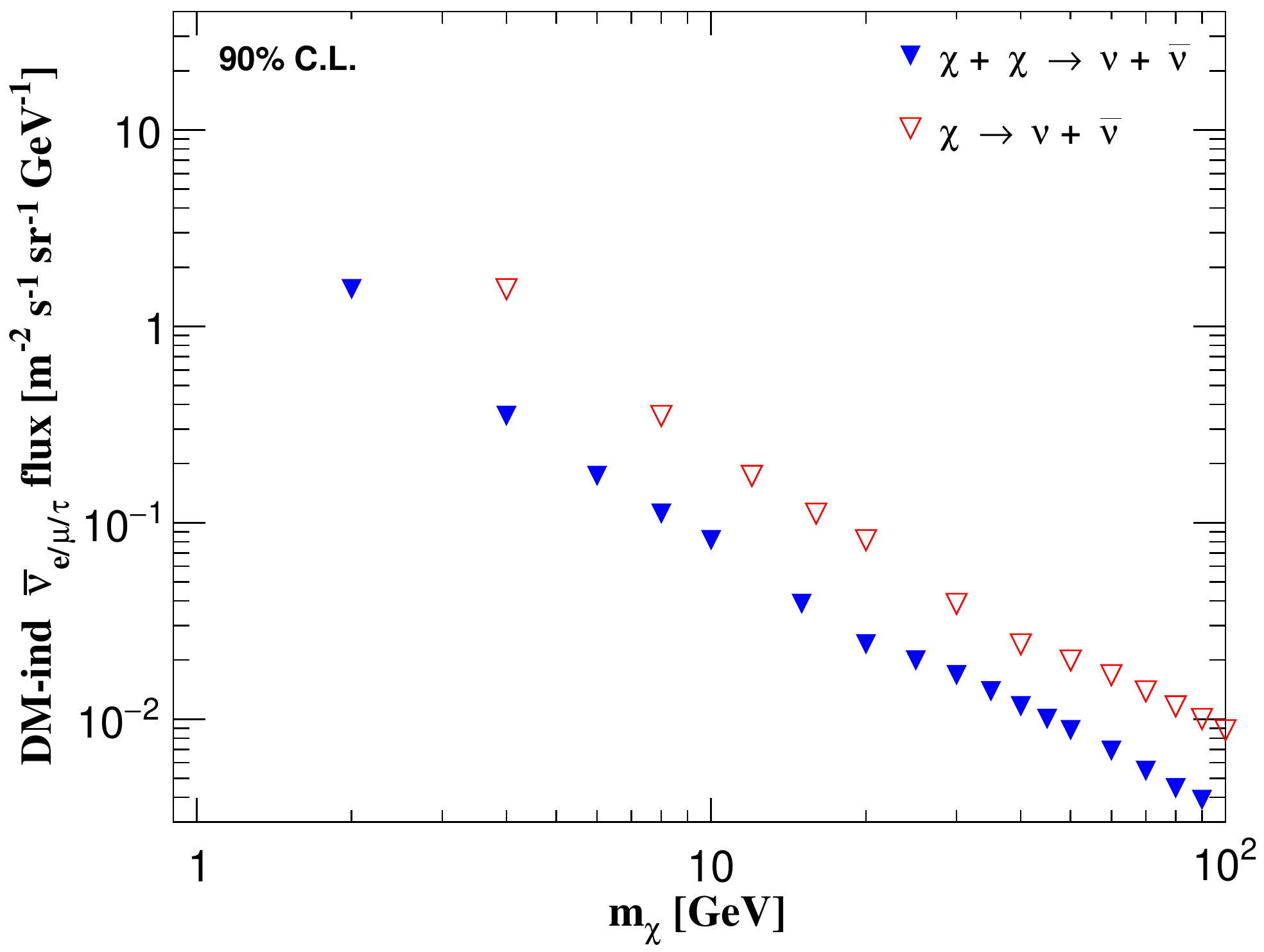}}
\mycaption{The limit on (a) $\nu_{e/\mu/\tau}$ and  (b) $\bar{\nu}_{e/\mu/\tau}$ flux produced from the dark matter in Milky Way galaxy
at 90$\%$ C.L. (1 d.o.f.) by 500 kt$\cdot$yr MagICAL detector. The blue filled and red empty 
triangles are for the annihilation and decay of dark matter particles respectively.}
\label{flux-bound}
\end{figure}
We can use the constraints on $\langle \sigma v \rangle$ (see section\,\ref{subsec:resultA}) and $\tau$ (see section\,\ref{subsec:resultB})
in Eqs.\,\ref{eq6} and \ref{eq10} respectively to place the upper bound on the neutrino and antineutrino flux from dark matter .
In Fig.\,\ref{flux-bound}(a), the blue filled triangles and red empty triangles depict the upper bounds on 
$\nu_{e}/\nu_{\mu}/\nu_{\tau}$ flux at 90$\%$ C.L. (1 d.o.f.) using the constraints on $\langle \sigma v \rangle$ (in case 
of annihilation) and $\tau$ (in case of decay) respectively.   
Fig.\,\ref{flux-bound}(b) shows the same for $\bar\nu_{e}/\bar\nu_{\mu}/\bar\nu_{\tau}$ flux.  
The mass ordering is taken as NO and the dark matter profile is assumed to be NFW. 
We can see from both the panels in Fig.\,\ref{flux-bound} that the limits on neutrino (left panel) 
and antineutrino (right panel) flux from both annihilation and decay 
improve as we increase the value of $m_\chi$.
We can understand this behavior in the following way.
We know that the atmospheric neutrino event rates which serve as background for annihilation 
and decay decrease as we go to higher neutrino energy. This can be clearly seen from Fig.\,\ref{fig4} and also Fig.\,\ref{fig6}. 
This is also true for atmospheric antineutrino events. Since, the atmospheric neutrino and antineutrino backgrounds get reduced
when we go from lower to higher $m_\chi$, we need less dark matter induced neutrino and antineutrino flux for both annihilation and decay 
to obtain the same confidence level in $\Delta\chi^2$ which is 2.71 at 90$\%$ C.L. (1 d.o.f).
Hence, we can place better constraints on the DM induced neutrino and anitneutrino flux as we move from lower to higher $m_\chi$ values.
Another feature that is emerging from both the panels in Fig.\,\ref{flux-bound} that we have better constraints on the neutrino and 
antineutrino flux obtained from the annihilation of dark matter as compare to its decay for a fixed $m_\chi$.  
We can also explain this feature in the following way. For a fixed value of $m_\chi$, the available energy of neutrino and antineutrino,
$E_{\nu/\bar\nu}$, is equal to $m_\chi$ for annihilation and $m_\chi$/2 for decay. Let us consider the case for $m_\chi\,=\,10$ GeV
in both the panels. In this case, the available neutrino/antineutrino energy for annihilation (decay) is 10 GeV (5 GeV).
Now, we already know that the background events induced by atmospheric neutrino and anitneutrino flux
are higher at 5 GeV (in case of decay) as compared to 10 GeV (in case of annihilation).
Therefore, for a fixed choice of $m_\chi$ value, we need higher neutrino and antineutrino flux 
from decaying DM as compare to annihilating DM to place the constraints at same confidence level.


\section{Conclusions}
\label{sec:conclusions}
We explore the prospects of detecting diffuse dark matter in the Milky Way galaxy at the proposed INO-MagICAL detector.
The future sensitivity of 500 kt$\cdot$yr MagICAL detector to constrain the  dark matter self-annihilation cross-section ($\langle\sigma v\rangle$)
and decay lifetime ($\tau$) for $\chi\chi\rightarrow\nu\bar\nu$ and $\chi\rightarrow\nu\bar\nu$ processes respectively are estimated.  
We find that MagICAL will be able to probe new parameter space for low mass dark matter.

Combining information from $\nu$ and $\bar{\nu}$ modes, the future limits on $\langle\sigma v\rangle$ and $\tau$ are 
$\leq$ 1.87 $\times\,10^{-24}$ cm$^{3}$ s$^{-1}$ and $\geq$ 4.8 $\times\, 10^{24}$ s respectively at 90$\%$ C.L. (1 d.o.f.)  for $m_\chi$ = 10 GeV assuming the 
NFW profile.  These limits will be novel and they will address many viable dark matter models.  
The limits for higher dark matter masses will also be competitive with other neutrino telescopes.  

We have also shown the bounds on $\langle\sigma v\rangle$
and $\tau$ with  $\nu$ and $\bar\nu$ data separately. This enables us to probe the same physics through the $\nu$ and $\bar\nu$
channels due to the charge identification capability of the MagICAL detector.   

Although, we have studied the processes $\chi\chi\rightarrow\nu\bar\nu$ and $\chi\rightarrow\nu\bar\nu$, other final states 
like $\mu^+ \mu^-$, $\tau^+ \tau^-$, $b \bar b$ are also possible.  The constraints on these channels obtained from the gamma-ray 
detectors are much stronger, and hence we do not consider them.  Since the analysis is done for the diffuse dark matter 
component of the Milky Way galaxy, the constraints on self-annihilation cross-section and decay lifetime are robust 
and conservative, and the constraints have mild dependence on the dark matter profile.
Besides new and novel methods in dark matter indirect detection physics\,\cite{Speckhard:2015eva,Powell:2016zbo}, it is 
imperative that we fully utilize the capabilities of new and upcoming detectors. Our work explores the capabilities of 
INO-MagICAL to search for dark matter, and we encourage the community to look into this signature in more detail.
\vspace{.4 cm}
\section{Acknowledgment}
A.K. would like to thank the INO project for financial support.  R.L. thanks KIPAC for financial help.  S.K.A. is supported by DST/INSPIRE Research Grant
No. IFA-PH-12, Department of Science and Technology, India.  We thank Amol Dighe, Pankaj Agrawal, Ajit Srivastava, and Tarak Thakore for useful discussions.  
\appendix
\section{Oscillation of DM induced neutrinos}
\label{app1}
The oscillation probability of neutrino from one flavor ($\alpha$) to another flavor ($\beta$) in vacuum
is given by
\begin{align}
 P_{\alpha\beta} &=\sum_{k=l=1}^{3}\,|U_{\alpha k}|^2\,|\,U_{\beta l}|^2\,\nonumber\\
                            &+\,2\,\sum_{l>k}\,Re(\,U_{\alpha k}\,U^{*}_{\beta k}\,U^{*}_{\alpha l}\,U_{\beta l})\,\cos(\Delta E_{lk}\,L)\nonumber\\
                            &-\,2\,\sum_{l>k}\,Re(\,U_{\alpha k}\,U^{*}_{\beta k}\,U^{*}_{\alpha l}\,U_{\beta l})\,\sin(\Delta E_{lk}\,L)\,,
 \label{pab}
\end{align}
where U as the 3 $\times$ 3 unitary PMNS matrix \cite{Pontecorvo:1957qd,Maki:1962mu,Pontecorvo:1967fh}.
When $L$ is very large, 2nd and 3rd terms in Eq.\,\ref{pab} get averaged out to zero due to very rapid oscillations, 
and give rise to the following expression  
\begin{equation}
P_{\alpha\beta} = \sum_{k=1}^{3}\,|U_{\alpha k}|^2\,|\,U_{\beta k}|^2\,.
\label{eq_prob_vac}
\end{equation}
We assume that the annihilation/decay of dark matter particles produce $\nu_{e}$, $\nu_{\mu}$, and $\nu_\tau$ 
in the ratio of 1:1:1 at the source. During their propagation through the astronomical distance from  
source to detector, they undergo vacuum oscillation. 
Imposing the unitary property of U in Eq.\,\ref{eq_prob_vac}, the ratio of 
neutrino flavors at the Earth surface remains 
1:1:1, and this is true irrespective of the values of oscillation parameters.

\bibliographystyle{JHEP}
\bibliography{reference-ical-dm.bib}
\end{document}